\documentclass[12pt, preprint,preprintnumbers,nofootinbib, groupedaddress,superscriptaddress,amsmath,amssymb]{revtex4}
\usepackage{graphicx}
\usepackage{dcolumn}
\usepackage{bm}
\usepackage{amssymb}
\usepackage{amsmath}
\usepackage{epsfig}    
\usepackage{color}
\usepackage{hhline}

\def\be{\begin{equation}}
\def\ee{\end{equation}}
\newcommand{\bea}{\begin{eqnarray}}
\newcommand{\eea}{\end{eqnarray}}
\newcommand{\nn}{\nonumber}

\numberwithin{equation}{section}

\begin{document}
{\begin{flushright}{APCTP Pre2022 - 002, KEK-TH-2397, KYUSHU-HET-236}\end{flushright}}

\title{Radiative neutrino masses from modular $A_4$ symmetry and supersymmetry breaking}

\author{Hajime Otsuka}
\email{otsuka.hajime@phys.kyushu-u.ac.jp}
\affiliation{KEK Theory Center, Institute of Particle and Nuclear Studies, 1-1 Oho, Tsukuba, Ibaraki 305-0801, Japan}
\affiliation{Department of Physics, Kyushu University, 744 Motooka, Nishi-ku, Fukuoka, 819-0395, Japan}

\author{Hiroshi Okada}
\email{hiroshi.okada@apctp.org}
\affiliation{Asia Pacific Center for Theoretical Physics (APCTP) - Headquarters San 31, Hyoja-dong,
Nam-gu, Pohang 790-784, Korea}
\affiliation{Department of Physics, Pohang University of Science and Technology, Pohang 37673, Republic of Korea}

\date{\today}

\begin{abstract}
We investigate a modular $A_4$ invariant  two-loop neutrino mass model in a supersymmetric framework,
where we introduce new fields as minimum as possible, expecting contributions of superpartners to the neutrino masses. 
We successfully reproduce the neutrino oscillation data thanks to the superpartner contributions in case of normal hierarchy, and predict several observables such as phases and neutrino masses, concentrating on three specific regions at nearby fixed points of modulus $\tau=i,\omega,i\times \infty$, where $\omega\equiv e^{2\pi i/3}$. 
{These points are statistically favored in flux compactifications of the string theory.}
We show several results in each the points by performing global $\chi^2$ analysis, and demonstrate benchmark points with the minimum $\chi^2$.
\end{abstract}
\maketitle
\newpage

\section{Introduction}
Active neutrino and dark matter (DM) candidate may be related each other, since their features are similar in view of electric neutrality and rather difficulty of detection.
In fact, the nature of neutrino is partially discovered by experiments, and DM is not directly found even though so many different kinds of experiments are on going.
From a theoretical point of view, it could suggest that we propose more imaginable ideas than the other three fermion sector in the standard model (SM).
One of the ideas is a radiative seesaw model~\cite{Ma:2006km}, in which the neutrino mass matrix is constructed at loop levels.
When DM runs in the loop, we might interpret the neutrinos indirectly interact with the SM Higgs and get the masses only through DM.
Hence, we might understand the origin of tiny neutrino masses.
In order to realize this kind of neutrino masses, we often need a symmetry such as $Z_2$ that could simultaneously stabilizes DM.
\footnote{In this framework, a leptonic DM is favored.}
This symmetry is usually introduced by hand. But, if it arises from the other symmetry, the model would be more attractive.
Also, this scenario could accommodate other phenomenologies such as lepton flavor violations (LFVs) due to not so small Yukawa couplings. 
It implies that the model can be within a low energy scale that might reach at current experiments, thus the testability is enhanced.
The neutrino mixing patterns, phases, and LFVs do depend on the structure of Yukawa matrix, but the SM does not provide any prescriptions to fix this structure.

Several years ago, a finite modular group has been applied to the lepton sector to predict the neutrino oscillation data in ref.~\cite{Feruglio:2017spp}.
\footnote{Charged-lepton and neutrino sectors have been discussed in ref.~\cite{deAdelhartToorop:2011re} by embedding subgroups of various finite modular flavor symmetries.}
Interestingly, this scenario does not require many flavons that are traditionally introduced to get a desired texture, and the group includes a new degree of freedom{: ``modular weight",} that comes from the modular group. If we apply this group to a radiative seesaw model, DM can be stable by assigning nonzero modular weight.
In this way, we might obtain predictions to the lepton sector, and a specific interaction pattern of LFVs could be tested by current experiments.
Triggering this paper, a lot of groups have applied to various phenomenologies such as quark, lepton, DM, and so on.
For example, 
the modular $A_4$ flavor symmetry has been discussed in refs.~\cite{Feruglio:2017spp, Criado:2018thu, Kobayashi:2018scp, Okada:2018yrn, Nomura:2019jxj, Okada:2019uoy, deAnda:2018ecu, Novichkov:2018yse, Nomura:2019yft, Okada:2019mjf,Ding:2019zxk, Nomura:2019lnr,Kobayashi:2019xvz,Asaka:2019vev,Zhang:2019ngf, Gui-JunDing:2019wap,Kobayashi:2019gtp,Nomura:2019xsb, Wang:2019xbo,Okada:2020dmb,Okada:2020rjb, Behera:2020lpd, Behera:2020sfe, Nomura:2020opk, Nomura:2020cog, Asaka:2020tmo, Okada:2020ukr, Nagao:2020snm, Okada:2020brs, Yao:2020qyy, Chen:2021zty, Kashav:2021zir, Okada:2021qdf, deMedeirosVarzielas:2021pug, Nomura:2021yjb, Hutauruk:2020xtk, Ding:2021eva, Nagao:2021rio, king, Okada:2021aoi, Nomura:2021pld, Kobayashi:2021pav, Dasgupta:2021ggp, Liu:2021gwa, Nomura:2022hxs}, 
$S_3$  in refs.~\cite{Kobayashi:2018vbk, Kobayashi:2018wkl, Kobayashi:2019rzp, Okada:2019xqk, Mishra:2020gxg, Du:2020ylx}, 
$S_4$  in refs.~\cite{Penedo:2018nmg, Novichkov:2018ovf, Kobayashi:2019mna, King:2019vhv, Okada:2019lzv, Criado:2019tzk,
Wang:2019ovr, Zhao:2021jxg, King:2021fhl, Ding:2021zbg, Zhang:2021olk, gui-jun, Nomura:2021ewm}, 
$A_5$ in refs.~\cite{Novichkov:2018nkm, Ding:2019xna,Criado:2019tzk},
double covering of $A_4$  in refs.~\cite{Liu:2019khw, Chen:2020udk, Li:2021buv}, 
double covering of $S_4$  in refs.~\cite{Novichkov:2020eep, Liu:2020akv},   and
double covering of $A_5$  in refs.~\cite{Wang:2020lxk, Yao:2020zml, Wang:2021mkw, Behera:2021eut}.
Other types of modular symmetries have also been proposed to understand masses, mixings, and phases of the standard model (SM) in refs.~\cite{deMedeirosVarzielas:2019cyj, Kobayashi:2018bff,Kikuchi:2020nxn, Almumin:2021fbk, Ding:2021iqp, Feruglio:2021dte, Kikuchi:2021ogn, Novichkov:2021evw, Kikuchi:2021yog, Novichkov:2022wvg}.~\footnote{Here, we provide useful review references for beginners~\cite{Altarelli:2010gt, Ishimori:2010au, Ishimori:2012zz, Hernandez:2012ra, King:2013eh, King:2014nza, King:2017guk, Petcov:2017ggy}.}
Different applications to physics such as dark matter and origin {of} CP are found in refs.~\cite{Kobayashi:2021ajl, Nomura:2019jxj, Nomura:2019yft, Nomura:2019lnr, Okada:2019lzv, Baur:2019iai, Kobayashi:2019uyt, Novichkov:2019sqv,Baur:2019kwi, Kobayashi:2020hoc, Kobayashi:2020uaj, Ishiguro:2020nuf, Ishiguro:2021ccl, Tanimoto:2021ehw}.
Mathematical {studies} such as possible correction from K\"ahler potential, systematic analysis of the fixed points, 
and moduli stabilization are discussed in refs.~\cite{Chen:2019ewa, deMedeirosVarzielas:2020kji, Ishiguro:2020tmo, Abe:2020vmv, Novichkov:2022wvg}.
Recently, the authors of ref. \cite{Kikuchi:2022txy} proposed a scenario to derive four-dimensional modular flavor symmetric models from higher-dimensional theory on extra-dimensional spaces with the modular symmetry. 
It constrains modular weights and representations of fields and modular couplings in the four-dimensional effective field theory. 
Higher-dimensional operators in the SM effective field theory are also constrained in the higher-dimensional theory, in particular, the string theory \cite{Kobayashi:2021uam}. Non-perturbative effects relevant to neutrino masses are studied in the context of modular symmetry anomaly \cite{Kikuchi:2022bkn}.
%
%

In this paper, we apply a modular $A_4$ symmetry into a two-loop induced neutrino model in refs.~\cite{Kajiyama:2013zla, Kajiyama:2013rla}, where we introduce new fields as minimum as possible, expecting contributions of superpartners to the neutrino masses.~\footnote{The neutrino mass scenario from the soft {supersymmetry (SUSY)-breaking} terms are proposed by ref.~\cite{Arkani-Hamed:2000oup}.} {We focus on the SUSY-breaking sector inducing modular symmetric soft SUSY-breaking terms. (See for the SUSY-breaking phenomenology in modular flavor models \cite{Du:2020ylx,Kobayashi:2021jqu}).} 
In fact, we successfully reproduce the neutrino oscillation data thanks to these contributions in case of normal hierarchy, and predict several observables such as phases and neutrino masses at nearby three fixed points of modulus $\tau=i,\omega,i\times \infty$, where $\omega\equiv e^{2\pi i/3}$. 
These fixed points are statistically favored in the flux compactification of Type IIB string theory \cite{Kobayashi:2021pav}.
We show several results in each the points by performing global $\chi^2$ analysis, and demonstrate {benchmark points} with the minimum $\chi^2$.

This paper is organized as follows.
In Sec.~\ref{sec:realization}, we review our model, giving superpotential and SUSY-breaking terms. Then, we formulate valid mass matrices for bosons and fermions that are needed to construct the neutrino mass matrix.
In Sec.~\ref{sec:num},  we show several predictions at nearby three fixed points via global $\chi^2$ analysis, and demonstrate benchmark points with  the minimum $\chi^2$.
Finally, we conclude and summarize our model in Sec.~\ref{sec:conclusion}, in which we briefly discuss the possibility of DM candidate. 
In Appendix A, we summarize formulas in the framework of modular $A_4$ symmetry.

\begin{center} 
\begin{table}[tb]
\begin{tabular}{|c||c|c|c|c||c|c|c|c|c||}\hline\hline  
&\multicolumn{9}{c||}{ Chiral superfields}   \\\hline
  & ~$\{{\hat{L}_{e}},{\hat{L}_{\mu}},{\hat{L}_{\tau}}\}$~& ~$\{\hat{e}^c,\hat{\mu}^c,\hat{\tau}^c\}$~ & ~$\{\hat{N}^c_{1},\hat{N}^c_{2}\}$~& ~$\hat{S}$ ~& ~$\hat{H}_1$~ & ~$\hat{H}_2$~& ~$\hat{\eta}_1$~ & ~$\hat{\eta}_2$~ & ~$\hat{\chi}$~
  \\\hline 
 $SU(2)_L$ & $\bm{2}$  & $\bm{1}$  & $\bm{1}$ & $\bm{1}$   & $\bm{2}$  & $\bm{2}$  & $\bm{2}$  & $\bm{2}$ & $\bm{1}$   \\\hline 
$U(1)_Y$ & $-\frac12$ & $1$ & $0$  & $0$ & $\frac12$ & $-\frac12$& $\frac12$ & $-\frac12$  & $0$     \\\hline
 $A_4$ & $\{1,1'',1'\}$ & $\{1,1',1''\}$ & $\{1,1'\}$ & $3$ & $1$ & $1$ & $1$ & $1$ & $1$      \\\hline
 $-k_I$ & ${-4}$ & $0$ & $-1$ & $0$ & $0$ & $0$ & $-3$ & $-3$& $-3$ \\\hline
\end{tabular}
\caption{Field contents of {matter chiral superfields} and their charge assignments under $SU(2)_L\times U(1)_Y\times A_{4}$ in the lepton and boson sector, where $k$ is the number of modular weight, and the quark sector is the same as the SM.}
\label{tab:fields}
\end{table}
\end{center}

\section{ Model} 
\label{sec:realization}
Here, we review our model in order to obtain the two-loop neutrino masses 
{in a similar mechanism discussed in the non-SUSY framework \cite{ Kajiyama:2013rla}.}
In addition to {the minimal supersymmetric SM (MSSM), we introduce matter superfields including} two right-handed neutral fermions $N^c_{1,2}$ and three left-handed neutral fermions $S$.
$N^c_{1,2}$ and $S$ respectively belong to $1,1'$ and $3$ under {the modular} $A_4$ group. {Here, the modular $A_4$ group is one of the finite modular subgroups of $SL(2,\mathbb{Z})$, parametrized by the modulus $\tau$.}
We also add {two superfields including} inert bosons $\eta_1$ and $\chi$ where these are true singlets under the $A_4$ group.
{Chiral superfields $\{\hat{H}_2, \hat{\eta}_2 \}$ including two bosons $\{H_2, \eta_2\}$  are just required in order to retain the holomorphic feature. Here and in what follows, we denote by $\tilde{\phi}$ their superpartners in matter superfields ${\hat \phi}$.}
Nonzero modular weights are imposed by $-k_{(L_e,L_\mu,L_\tau)}=-4$, $k_{N^c_{1,2}}=-1$, $-k_{(\eta_{1,2},\chi)}=-3$. 
All the fields and their assignments are summarized in Table~\ref{tab:fields}.
Under these symmetries, one writes renormalizable superpotential as follows:
\begin{align}
&{\cal W} =
y_e   \hat{e}^c \hat{L}_{e} \hat{H}_2 + y_\mu   \hat{\mu}^c \hat{L}_{\mu} \hat{H}_2 +y_\tau  \hat{\tau}^c \hat{L}_{\tau} \hat{H}_2  
\nn\\
&+a_\eta Y^{(8)}_1   \hat{N}^c_{1} \hat{L}_{e} \hat{\eta}_1
+b_\eta Y^{(8)}_{1'} \hat{N}^c_{1} \hat{L}_{\mu} \hat{\eta}_1 
+d_\eta Y^{(8)}_{1''} \hat{N}^c_{1} \hat{L}_{\tau} \hat{\eta}_1 
+c_\eta Y^{(8)}_{1} \hat{N}^c_{2} \hat{L}_{\mu}  \hat{\eta}_1 
+d_\eta Y^{(8)}_{1'} \hat{N}^c_{2} \hat{L}_{\tau} \hat{\eta}_1 
+e_\eta Y^{(8)}_{1''} \hat{N}^c_{2} \hat{L}_{e} \hat{\eta}_1 
\nn\\
&+{\alpha_{NS}} \hat{N}^c_{1}  (y'_1 {\hat{S}_{1}} + y'_2 {\hat{S}_{2}}+ y'_3 {\hat{S}_{3}}) \hat{\chi}
 +{\beta_{NS}} \hat{N}^c_{2} (y'_2 {\hat{S}_{2}} + y'_1 {\hat{S}_{3}}+ y'_3 {\hat{S}_{1}})  \hat{\chi}
\nn\\
&+ M_{0} ( \hat{S}_{1} \hat{S}_{1} + \hat{S}_{2} \hat{S}_{3}  +  \hat{S}_{3} \hat{S}_{2}) 
\label{eq:sp-lep}
\\
&+\mu_H \hat{H}_1 \hat{H}_2+\mu_\eta Y^{(6)}_1 \hat{\eta}_1 \hat{\eta}_2+\mu_\chi Y^{(6)}_1 \hat{\chi} \hat{\chi} 
 + a Y^{(6)}_1 \hat{H}_1 \hat{\eta}_2 \hat{\chi} + b Y^{(6)}_1 \hat{H}_2 \hat{\eta}_1 \hat{\chi}, 
\nn
\end{align}
where R-parity is implicitly imposed in the above superpotential, $Y^{(2)}_3\equiv(y_1,y_2,y_3)^T$ is $A_4$ triplet with modular weight $2$, $Y^{(4)}_3\equiv(y'_1,y'_2,y'_3)^T$ is $A_4$ triplet with modular weight $4$, and $Y^{(8)}_{1,1',1''}$ are $A_4$ singlets with modular weight $8$.~\footnote{The concrete expressions of modular Yukawas are summarized in Appendix A.} The charged-lepton mass eigenvalues are given by the first term as $m_{e,\mu,\tau}\equiv y_{e,\mu,\tau} v_2/\sqrt2$ after the spontaneous symmetry breaking, in the above equation that is diagonal.
Here, vacuum expectation values of $H_{1,2}$ denote $[0,v_1/\sqrt2]^T$ and $[v_2/\sqrt2,0]^T$, respectively. 
Therefore, we do not need to consider the mixing of charged-lepton sector.

Valid soft SUSY-breaking terms to construct the neutrino mass matrix are found as follows: 
\begin{align}
&-{\cal L}_{\rm soft} = \mu_{BH}^2 H_1 H_2 + \mu_{B\eta}^2 Y^{(6)}_1 \eta_1\eta_2+ \mu_{B\chi}^2 Y^{(6)}_1 \chi\chi +
A_a Y^{(6)}_1 H_1\eta_2 \chi+ A_b Y^{(6)}_1 H_2\eta_1 \chi\nn\\
&+m^2_{\tilde S} \left(|\tilde S_1|^2+|\tilde S_2|^2+|\tilde S_3|^2\right)
+m^2_{\tilde N^c_1}|\tilde N^c_1|^2
+m^2_{\tilde N^c_2}|\tilde N^c_2|^2
+m^2_{H_1}|H_1|^2+m^2_{H_2}|H_2|^2
+m^2_{\eta_1}|\eta_1|^2+m^2_{\eta_2}|\eta_2|^2\nn\\
&
+m^2_{\chi}|\chi|^2
+A_{a_\eta} Y^{(4)}_1 \tilde N^c_1\tilde L_e \eta_1
+A_{b_\eta} Y^{(4)}_{1'} \tilde N^c_1\tilde L_\mu \eta_1
+A_{c_\eta} Y^{(4)}_{1} \tilde N^c_2\tilde L_\mu \eta_1
+A_{d_\eta} Y^{(4)}_{1'} \tilde N^c_2\tilde L_\tau \eta_1 \\
&+A_{\alpha_{NS}} \tilde N^c_{1}  (y'_1 {\tilde  S_{1}} + y'_2 {\tilde S_{2}}+ y'_3 {\tilde S_{3}}) \chi
 +A_{\beta_{NS}} \tilde N^c_{2} (y'_2 {\tilde S_{2}} + y'_1 {\tilde S_{3}}+ y'_3 {\tilde S_{1}})  \chi
+\mu_{\tilde SB}^2(\tilde S^2_1+2\tilde S_2\tilde S_3)
+ {\rm h.c.},\nn \label{eq:pot}
\end{align}
where all fields are bosons, and $m^2_{\tilde N^c_{1,2}}$, $m^2_{\eta_{1,2}}$, $m^2_{\chi}$
includes the invariant coefficients $1/(\tau^*-\tau)^{k_{\tilde N^c_{1,2},\eta_{1,2}}}$. 
{The $A$-terms $\{A_{a_\eta}, A_{b_\eta}, A_{c_\eta}, A_{d_\eta} \}$ are irrelevant for the following discussion on  neutrino mass matrix.}
Here, we suppose the modular symmetric soft terms which will be generated by contributions of supersymmetry breaking fields $X$ including the modulus $\tau$. 
It is remarkable that these soft terms are invariant under the modular symmetries when the $X$ correspond to the modulus $\tau$. Indeed, the modulus field and its $F$-term are the same representations under the modular symmetry, as pointed out in ref. \cite{Kobayashi:2021uam}.\footnote{{This discussion is applicable to the multi moduli case realized in the higher-dimensional theory on Calabi-Yau manifolds \cite{Ishiguro:2020nuf,Ishiguro:2021ccl}.}} 
When we denote the $K_{\rm mod}$ and $K_{\rm matter}$ for the modulus and matter K\"ahler potentials, respectively,\footnote{The matter K\"ahler metric is assumed to be a diagonal form.} the explicit form of soft scalar masses $m_{\phi_{i}}^2$ and the $A$-terms {in the canonical normalization} are written as \cite{Kaplunovsky:1993rd}:
\begin{align}
    m_{\phi_{i}}^2 &= m_{3/2}^2 - \sum_X |F^X|^2 \partial_X\partial_{\bar{X}}\ln \partial_{\phi_{i}} \partial_{\bar{\phi}_i}K_{\rm matter},
    \nonumber\\
    A_{ijk} &= A_{i} + A_{j} +A_{k} - \sum_X \frac{F^X}{y_{ijk}} \partial_X (y_{ijk}),
\end{align}
with
\begin{align}
    A_i = \sum_X F^X \partial_X \ln \left(e^{-K_{\rm mod}} \partial_{\phi_{i}} \partial_{\bar{\phi}_i}K_{\rm matter}\right),
\end{align}
where $m_{3/2}$ is the gravitino mass and {$y_{ijk}$ denote the Yukawa couplings of fields. 
Since the modulus $\tau$ does not appear in gauge kinetic functions at the tree-level in the context of Type IIB string theory, we assume that the gaugino masses are generated by $F$-terms of other moduli  (see for the stabilization of moduli fields, e.g., refs. \cite{Kobayashi:2019xvz,Kobayashi:2020uaj,Kobayashi:2020hoc,Ishiguro:2020tmo,Ishiguro:2020nuf,Novichkov:2022wvg}). In this paper, we randomly search for soft terms to study the structure of neutrino masses and mixing angles without specifying the SUSY-breaking sector.}

 \subsubsection{Inert boson and fermion mixings}
Inert bosons $\chi$, $\eta_1$, and $\eta_2$ mix each other through the soft {SUSY-breaking} terms of $A_{a,b}$ and $\mu_{B\eta}$, after the spontaneous electroweak symmetry breaking.
Here, we suppose to be $\mu_{B\eta},\ A_a<<A_b$ for simplicity, then the mixing dominantly comes from $\chi$ and $\eta_1$ only. 
This assumption does not affect the structure of the neutrino mass matrix.
Then the mass eigenstate is defined by
\begin{align}
\left[\begin{array}{c}
\chi_{R,I} \\ 
\eta_{1_{R,I}}  \\ 
\end{array}\right]=
\left[\begin{array}{cc}
c_{\theta_{R,I}} & -s_{\theta_{R,I}}  \\ 
s_{\theta_{R,I}} & c_{\theta_{R,I}}   \\ 
\end{array}\right]
\left[\begin{array}{c}
\xi_{1_{R,I}}   \\ 
\xi_{2_{R,I}}    \\ 
\end{array}\right],
\end{align}
 where $c_{\theta_{R,I}}, s_{\theta_{R,I}}$ are respectively the shorthand notations of $\sin\theta_{R,I}$ and $\cos\theta_{R,I}$,
 and $\xi_{1,2}$ are mass eigenstates for $\chi,\eta_1$ and their mass eigenvalues are denoted by $m_{i_{R,I}},\ (i=1,2)$. Notice that the mixing angle $\theta$ simultaneously diagonalizes the mass matrix of real and imaginary part.
 
These superpartners of $ \chi$ and $ \eta$; $\tilde \chi$ and $\tilde \eta$, mix each other via $b$, and its mass matrix is given by
 \begin{align}
 {\cal M}_{\tilde\chi\tilde\eta}=Y^{(6)}_{1}
\left[\begin{array}{cc}
m_1 &m_2  \\ 
m_2 & 0   \\ 
\end{array}\right],
\end{align}
where $m_1\equiv \mu_{\chi}$, $m_2\equiv \frac{v_2 b}{2\sqrt2}$. Then, the above matrix is diagonalized by
a unitary matrix: 
 \begin{align}
&O  {\cal M}_{\tilde\chi\tilde\eta} O^T={\rm diag}(m_{\tilde\xi_1},m_{\tilde\xi_2}),\\
&O=
\left[\begin{array}{cc}
i&0 \\ 
0 &1  \\ 
\end{array}\right]
\left[\begin{array}{cc}
c_{\tilde\theta} & -s_{\tilde\theta}  \\ 
s_{\tilde\theta} & c_{\tilde\theta}   \\ 
\end{array}\right],\quad
\tan\tilde\theta = \frac{m_1+\sqrt{m_1^2+4m_2^2}}{2m_2},
\end{align}
where 
$m_{\tilde\xi_1}\equiv \frac12(\sqrt{m_1^2+4m_2^2}-m_1)$, $m_{\tilde\xi_2}\equiv \frac12(\sqrt{m_1^2+4m_2^2}+m_1)$, 
and $c_{\tilde\theta}, s_{\tilde\theta}$ are respectively the shorthand notations of $\sin\tilde\theta$ and $\cos\tilde\theta$.
Similar to the boson sector, the mass eigenstate is defined by
\begin{align}
\left[\begin{array}{c}
\tilde\chi \\ 
\tilde\eta_1\\ 
\end{array}\right]=
\left[\begin{array}{cc}
i c_{\tilde\theta} & s_{\tilde\theta}  \\ 
-i s_{\tilde\theta} & c_{\tilde\theta}   \\ 
\end{array}\right]
\left[\begin{array}{c}
\tilde\xi_{1}   \\ 
\tilde\xi_{2}    \\ 
\end{array}\right].
\end{align}

\subsubsection{Neutral fermion mass matrix of $S$}
The mass matrix of  $S$ is found to be
\begin{align}
{\cal M}_S&=
M_0
\left[\begin{array}{ccc}
1 &0 &0 \\ 
0& 0 & 1  \\ 
0& 1 & 0    \\ 
\end{array}\right]. \label{eq:mn}
\end{align}
Since ${\cal M}_S{\cal M}_S^\dag={\cal M}_S^\dag{\cal M}_S\sim 1_{3\times3}$, 
the mixing matrix $V_S=1$. Here, we define $S =V_S^T \psi =\psi $,
where $\psi$ is the mass eigenstate of $S$.

\subsubsection{Boson mass matrix of the superpartner $S$: $\tilde S$}
Since $\tilde S_i$ consists of real and imaginary scalars, 
we redefine them to be $\tilde S_i\equiv (\tilde s_{R_i}+i\tilde s_{I_i})/\sqrt2$. Then, we explicitly write the mass matrices as follows:
\begin{align}
{\cal M}^2_{\tilde S_R}&=
\left[\begin{array}{ccc}
|M_0|^2+m^2_{\tilde S}+\mu_{\tilde SB}^2 &0 &0 \\ 
0& 4 |M_0|^2+m^2_{\tilde S} & 2\mu_{\tilde SB}^2  \\ 
0& 2\mu_{\tilde SB}^2 & 4 |M_0|^2+m^2_{\tilde S}    \\ 
\end{array}\right], \label{eq:mbsr}\\
{\cal M}^2_{\tilde S_I}&=
\left[\begin{array}{ccc}
|M_0|^2+m^2_{\tilde S} -\mu_{\tilde SB}^2 &0 &0 \\ 
0& 4 |M_0|^2+m^2_{\tilde S} & -2\mu_{\tilde SB}^2  \\ 
0& -2\mu_{\tilde SB}^2 & 4 |M_0|^2+m^2_{\tilde S}    \\ 
\end{array}\right]. \label{eq:mbsr}
\end{align}
Both of the above mass matrices are diagonalized by an orthogonal matrix $O_{\tilde S}$ as follows:
\begin{align}
D^2_{\tilde S_R}&=
{\rm diag}[|M_0|^2+m^2_{\tilde S}+\mu_{\tilde SB}^2 ,4 |M_0|^2+m^2_{\tilde S}-2\mu_{\tilde SB}^2 ,4 |M_0|^2+m^2_{\tilde S}+ 2\mu_{\tilde SB}^2]
= O_{\tilde S} {\cal M}^2_{\tilde S_R} O_{\tilde S}^T, \\
D^2_{\tilde S_I}&=
{\rm diag}[|M_0|^2+m^2_{\tilde S}-\mu_{\tilde SB}^2 ,4 |M_0|^2+m^2_{\tilde S}+2\mu_{\tilde SB}^2 ,4 |M_0|^2+m^2_{\tilde S}- 2\mu_{\tilde SB}^2]
= O_{\tilde S} {\cal M}^2_{\tilde S_R} O_{\tilde S}^T, \\
&O_{\tilde S}
=
\left[\begin{array}{ccc}
1 &0 &0 \\ 
0& 1/\sqrt2 & -1/\sqrt2  \\ 
0& 1/\sqrt2 &  1/\sqrt2    \\ 
\end{array}\right].
\end{align}
Then, we find $\tilde s_{R,I} =O_{\tilde S}^T \tilde S_{R,I} $, where $\tilde s_{R,I}$ is the mass eigenstate of $\tilde S_{R,I}$.

\subsubsection{Neutral fermion mass matrix of $N^c$}
The mass of $N^c$ is induced at one-loop level, running $\psi$ and $\chi$.
The valid Lagrangian in terms of mass eigenstate of $S$ is found as follows:
\footnote{We neglect a contribution from gauginos to the neutrino masses~\cite{Megrelidze:2016fcs}, assuming the masses are much heavier than the masses of $N^c$, $\tilde N^c$, $\xi_{1_{R,I}}$, and , $\xi_{2_{R,I}}$. {This assumption will be justified when gaugino masses are generated by moduli fields without $\tau$.}}
\begin{align}
-&{\cal L} = N^c  \frac{ Y_{NS} }{\sqrt2}  \psi (c_{\theta_R} \xi_{1_R}-s_{\theta_R} \xi_{2_R})
- N^c  \frac{ Y_{NS}}{\sqrt2}  \psi (c_{\theta_I} \xi_{1_I}-s_{\theta_I} \xi_{2_I})
+{\rm h.c.},
\nn\\
 Y_{NS}&=
\left[\begin{array}{cc}
{\alpha_{NS}} &0  \\ 
0& {\beta_{NS}}   \\ 
\end{array}\right]
\left[\begin{array}{ccc}
y'_1 & y'_3 & y'_2 \\ 
y'_2 & y'_1 & y'_3 \\ 
\end{array}\right].
\end{align}
Then, the mass matrix of $N^c$ is derived as follows:
\begin{align}
&M_{N^c} = -  \frac{M_0}{(4\pi)^2}Y_{NS}  Y_{NS}^T\nn\\
&\times 
\left[
2F_0(M_0,m^2_{1_R},m^2_{1_I})(c^2_{\theta_R}-c^2_{\theta_I})
+
2F_0(M_0,m^2_{2_R},m^2_{2_I})(s^2_{\theta_R}-s^2_{\theta_I})
\right.\\
&\left.
+F_I(M_0,m^2_{1_R},m^2_{1_I})(c^2_{\theta_I}m^2_{1_R} -c^2_{\theta_R}m^2_{1_I})
+
F_I(M_0,m^2_{2_R},m^2_{2_I})(s^2_{\theta_I}m^2_{2_R}-s^2_{\theta_R}m^2_{2_I})
\right], \nn\\
&F_0(M_0,m_1,m_2)\equiv\int[dx]_3\ln[x M^2_0 +y m_1^2+z m_2^2],\quad
F_{I}(M_0, m_1,m_2)\equiv
\int\frac{[dx]_3}{x M^2_0 +y m_1^2+z m_2^2},
\end{align}
where $\int[dx]_3\equiv \int_0^1dxdydz\delta(1-x-y-z)$ and $m_{i_{R,I}}$ is the mass of $\xi_{i_{R,I}},\ (i=1,2)$.
%
Similar to $S$, we find $N^c=V_N^T \psi^c_N$, where $\psi^c_N$ is the mass eigenstate of $N^c$ and $D_N= V_N M_N V_N^T$.

\subsubsection{Boson mass matrix of the superpartner $N^c$: $\tilde N^c$}
The mass matrix of $\tilde N^c$ is also induced at one-loop level, running several fields.
The valid Lagrangian is given by 
\begin{align}
-{\cal L} &= \tilde N^c Y_{NS}  \psi (i c_{\tilde\theta} \tilde\xi_{1}+s_{\tilde\theta} \tilde\xi_{2})\nn\\
&+\tilde N^c  
\frac{ {\cal A} Y_{NS} O_{\tilde S}^T}{\sqrt2}  (\tilde s_R + i\tilde s_I)
\left[
(c_{\theta_R} \xi_{1_R}-s_{\theta_R} \xi_{2_R})
+
i(c_{\theta_I} \xi_{1_I}-s_{\theta_I} \xi_{2_I})
\right]
+{\rm h.c.} \label{eq:ncb} \\
&= \tilde N^c Y_{NS}   \psi (i c_{\tilde\theta} \tilde\xi_{1}+s_{\tilde\theta} \tilde\xi_{2})\nn\\
&
+c_\theta \tilde N^c  G (\tilde s_R + i\tilde s_I)(\xi_{1_R}+ i \xi_{1_I})
-s_\theta \tilde N^c  G (\tilde s_R + i\tilde s_I)(\xi_{2_R}+ i \xi_{2_I})
+{\rm h.c.},
\nn\\
{\cal A} Y_{NS} 
&=Y^{(4)}_1
\left[\begin{array}{cc}
A_{\alpha_{NS}} &0  \\ 
0& A_{\beta_{NS}}   \\ 
\end{array}\right]
\left[\begin{array}{ccc}
y'_1 & y'_3 & y'_2 \\ 
y'_2 & y'_1 & y'_3 \\ 
\end{array}\right],\quad
G\equiv {\cal A} Y_{NS}  O_{\tilde S}^T.
\end{align}
Then, the mass matrix $\tilde N^c$ is found as
\begin{align}
(M_{\tilde N^c}^2)_{ij}&= 
4 \frac{(Y_{\tilde N})_{ij} }{(4\pi)^2} 
\left[2(-c_{\tilde\theta}^2 m_{\tilde\xi_1} +s_{\tilde\theta}^2m_{\tilde\xi_2}) F_0(M_0,m_{\tilde\xi_1},m_{\tilde\xi_2})
+m_{\tilde\xi_1}m_{\tilde\xi_2}(c_{\tilde\theta}^2 m_{\tilde\xi_2} +s_{\tilde\theta}^2m_{\tilde\xi_1})
F_{I}(M_0, m_{\tilde\xi_1},m_{\tilde\xi_2})
\right]
\nn\\
&- \frac{G_{ia}G^T_{aj} }{(4\pi)^2}
\left(
2(c^2_{\theta_R} - c^2_{\theta_I}) [F_0(D_{\tilde S_{R_a}},m_{\xi_{1_R}},m_{\xi_{1_I}})
+F_0(D_{\tilde S_{I_a}},m_{\xi_{1_R}},m_{\xi_{1_I}})]
\right.\nn\\&\left.
+
2(s^2_{\theta_R} - s^2_{\theta_I}) [F_0(D_{\tilde S_{R_a}},m_{\xi_{2_R}},m_{\xi_{2_I}})
+F_0(D_{\tilde S_{I_a}},m_{\xi_{2_R}},m_{\xi_{2_I}})]\right.\nn\\
&\left.+
(c^2_{\theta_I}m^2_{1_R} -c^2_{\theta_R}m^2_{1_I}) [F_1(D_{\tilde S_{R_a}},m_{\xi_{1_R}},m_{\xi_{1_I}})
+F_1(D_{\tilde S_{I_a}},m_{\xi_{1_R}},m_{\xi_{1_I}})]
\right.\nn\\&\left.
+
(s^2_{\theta_I}m^2_{2_R}-s^2_{\theta_R}m^2_{2_I})[F_1(D_{\tilde S_{R_a}},m_{\xi_{2_R}},m_{\xi_{2_I}})
+F_1(D_{\tilde S_{I_a}},m_{\xi_{2_R}},m_{\xi_{2_I}})]
\right),\\
Y_{\tilde N}&=
\left[\begin{array}{cc}
{\alpha_{NS}^2}(y'^2_1+2y'_2y'_3) &\alpha_{NS}\beta_{NS}(y'^2_2+2y'_1y'_3)  \\ 
\alpha_{NS}\beta_{NS}(y'^2_2+2y'_1y'_3) &{\beta_{NS}^2}(y'^2_3+2y'_1y'_2)   \\ 
\end{array}\right].
\end{align}
%
Similar to $N^c$, we find $\tilde N^c=O_{\tilde N}^T \tilde n^c$, where $\tilde n^c$ is the mass eigenstate of $\tilde N^c$ and $D^2_{\tilde N}= O_{\tilde N} M_{\tilde N^c}^2 O_{\tilde N}^T$.

 \subsubsection{Mass matrix of $m_\nu$}
Now the active neutrino mass matrix $m_\nu$ is induced at one-loop level via the following Lagrangian in terms of mass eigenstate:
\begin{align}
-&{\cal L} =  
\frac1{\sqrt2}(\psi^c_N)_a (V_N)_{ai} (Y_\eta)_{ij} 
\nu_j(s_{\theta_R}\xi_{1_R}+c_{\theta_R}\xi_{2_R})
+
\frac {i}{\sqrt2}(\psi^c_N)_a (V_N)_{ai} (Y_\eta)_{ij} 
\nu_j(s_{\theta_I}\xi_{1_I}+c_{\theta_I}\xi_{2_I})\nn\\
&+
(\tilde n^c)_\alpha (O_{\tilde N})_{\alpha\beta}(Y_\eta)_{\beta j} \nu_j 
(-i s_{\tilde\theta}\tilde\xi_1 + c_{\tilde\theta}\tilde\xi_2)
+{\rm h.c.},\\
Y_\eta&=\frac{1}{\sqrt2}
\left[\begin{array}{ccc}
a_\eta Y^{(8)}_1 & b_\eta Y^{(8)}_{1'} & e_\eta  Y^{(8)}_{1''} \\ 
 f_\eta  Y^{(8)}_{1''} & c_\eta Y^{(8)}_1 & d_\eta Y^{(8)}_{1'}    \\ 
\end{array}\right] ,
\end{align}
Then, the neutrino mass matrix is given by
\begin{align}
m_\nu &=- \frac{1}{2(4\pi)^2}
(Y^T_\eta)_{i\alpha} (V^T_N)_{\alpha a} D_{N_a} (V_N)_{a\beta} (Y_\eta)_{\beta j}\times
\nn\\
&\left[
s^2_{\theta_R} f(m_{\xi_{1_R}},D_{N_a}) +
c^2_{\theta_R} f(m_{\xi_{2_R}},D_{N_a})
-s^2_{\theta_I} f(m_{\xi_{1_I}},D_{N_a})
-c^2_{\theta_I} f(m_{\xi_{2_I}},D_{N_a})
\right]\nn\\
&- \frac{\mu_\chi Y^{(6)}_1}{(4\pi)^2} m_{\tilde\xi_1} m_{\tilde\xi_2}
(Y_\eta^T)_{i\beta}(O_{\tilde N}^T)_{\beta\alpha}
F_{I}(D_{\tilde N_\alpha}, m_{\tilde\xi_1},m_{\tilde\xi_2})
(O_{\tilde N})_{\alpha\beta'}  (Y_\eta)_{\beta'j} 
,
\label{eq:mnu}\\
f(m_1,m_2)&=\int_0^1\ln\left[
x\left(\frac{m_1^2}{m_2^2}-1\right)+1
\right],
\end{align}
$m_\nu$ is diagonalized by a unitary matrix $U_\nu$ as follows: $U_\nu m_\nu U_\nu^T\equiv {\rm diag}[m_1,m_2,m_3]$, where $\sum_{i=1,2,3} m_{i}\lesssim 0.12$ eV is given by the recent cosmological data~\cite{Aghanim:2018eyx}.
Since the mixing matrix for charged-lepton is three by three unit matrix, one finds $U_\nu=U_{PMNS}$. It is remarkable that the second term in eq. (\ref{eq:mnu}) represents the loop effects of superpartners, and 
these contributions play an important role of obtaining the rank-3 neutrino mass matrix. 
Atmospheric mass square difference $\Delta m^2_{\rm atm}$ is written by
\begin{align}
\Delta m^2_{\rm atm} =  m_3^2- m_1^2\quad {\rm Normal\ Hierarchy\ (NH)},\\
\Delta m^2_{\rm atm} =  m_2^2- m_3^2\quad {\rm Inverted\ Hierarchy\ (IH)}.
\end{align}
While the solar mass difference square is found as
\begin{align}
&\Delta m_{\rm sol}^2= m_{2}^2- m_{1}^2.
\end{align}
Each of mixing angle is given in terms of the component of $U_{PMNS}$ as follows:
\begin{align}
\sin^2\theta_{13}=|(U_{PMNS})_{13}|^2,\quad 
\sin^2\theta_{23}=\frac{|(U_{PMNS})_{23}|^2}{1-|(U_{PMNS})_{13}|^2},\quad 
\sin^2\theta_{12}=\frac{|(U_{PMNS})_{12}|^2}{1-|(U_{PMNS})_{13}|^2}.
\end{align}
The effective mass for the neutrinoless double beta decay is depicted by
\begin{align}
\langle m_{ee}\rangle= |  m_1 \cos^2\theta_{12} \cos^2\theta_{13}+  m_2 \sin^2\theta_{12} \cos^2\theta_{13}e^{i\alpha_{21}}
+  m_3\sin^2\theta_{13}e^{i(\alpha_{31}-2\delta_{\rm CP})}|,
\end{align}
where $\delta_{\rm CP},(\alpha_{21},\alpha_{31})$ are respectively Dirac and two Majorana CP phases appearing in the $U_{PMNS}$ matrix.
$\langle m_{ee}\rangle$ could be tested by KamLAND-Zen in future~\cite{KamLAND-Zen:2016pfg}.

\section{Numerical analysis}
\label{sec:num}
In this section, we show numerical $\Delta \chi^2$ analysis at nearby three fixed points, employing the five reliable experimental data; $\Delta m^2_{\rm atm}, \Delta m_{\rm sol}^2, \sin^2\theta_{13},\sin^2\theta_{23},\sin^2\theta_{12}$ in ref.~\cite{Esteban:2018azc}.
Notice here that we consider CP phases $\delta_{\rm CP},\alpha_{21},\alpha_{31}$ as predictive values, and three charged-lepton masses are supposed to be fitted by the best fit values.
In case of IH, we would not find any allowed region within $5\sigma$. 
Thus, we focus on the case of NH only. {The dimensionful} input parameters are randomly selected by the range of [$10^{2}-10^7$] GeV except for $\mu_\chi$ with [$10^{-5}-10$] GeV, 
while the dimensionless ones [$10^{-10}-10^{-1}$] {except for $\tau$}. 
We work on three fixed points in the fundamental region of $\tau$, and all the input parameters are supposed to be real. Therefore, the origin of CP comes from $\tau$.

\subsection{$\tau=i$} 
\begin{figure}[tb!]\begin{center}
\includegraphics[width=80mm]{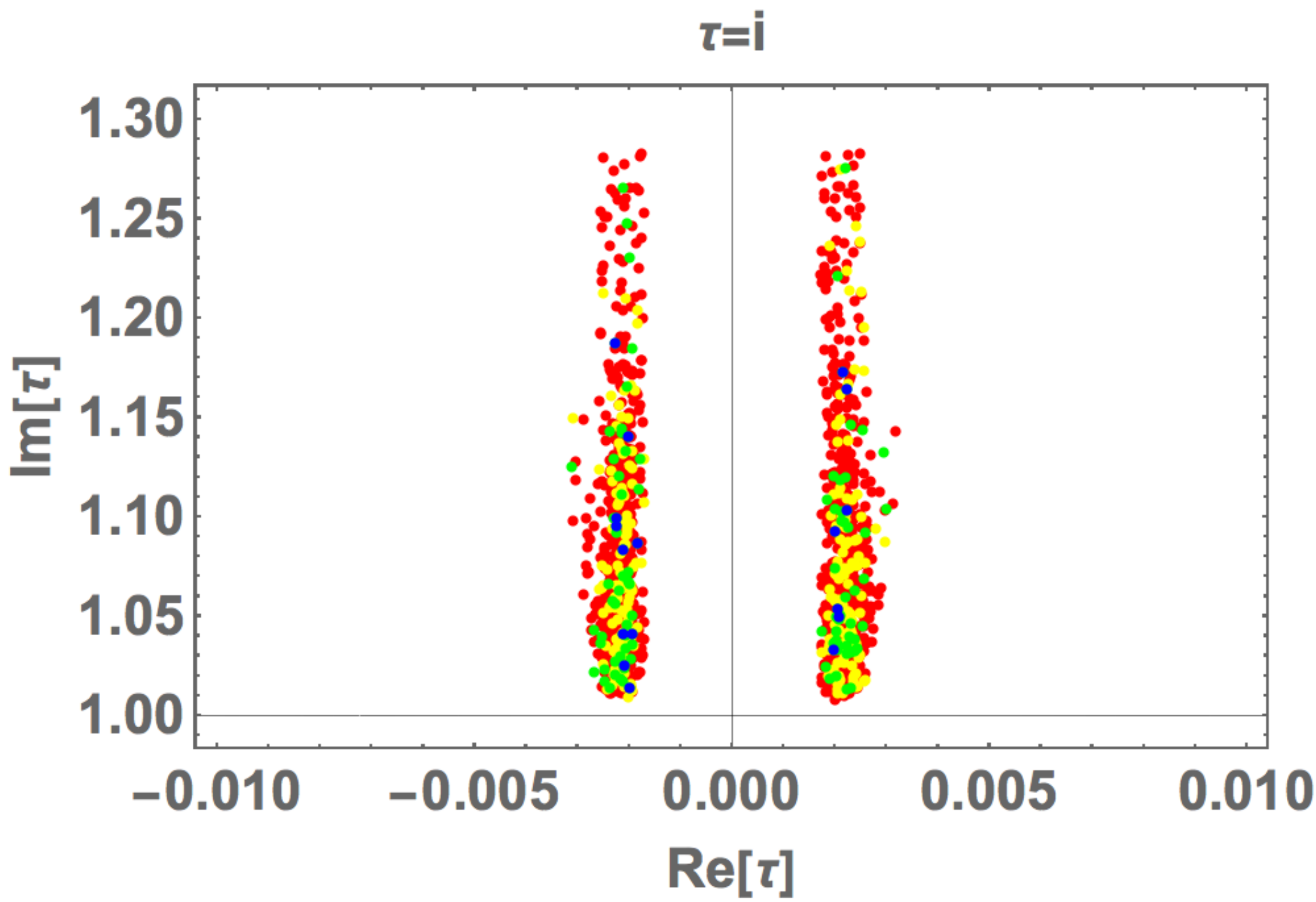} 
 \includegraphics[width=80mm]{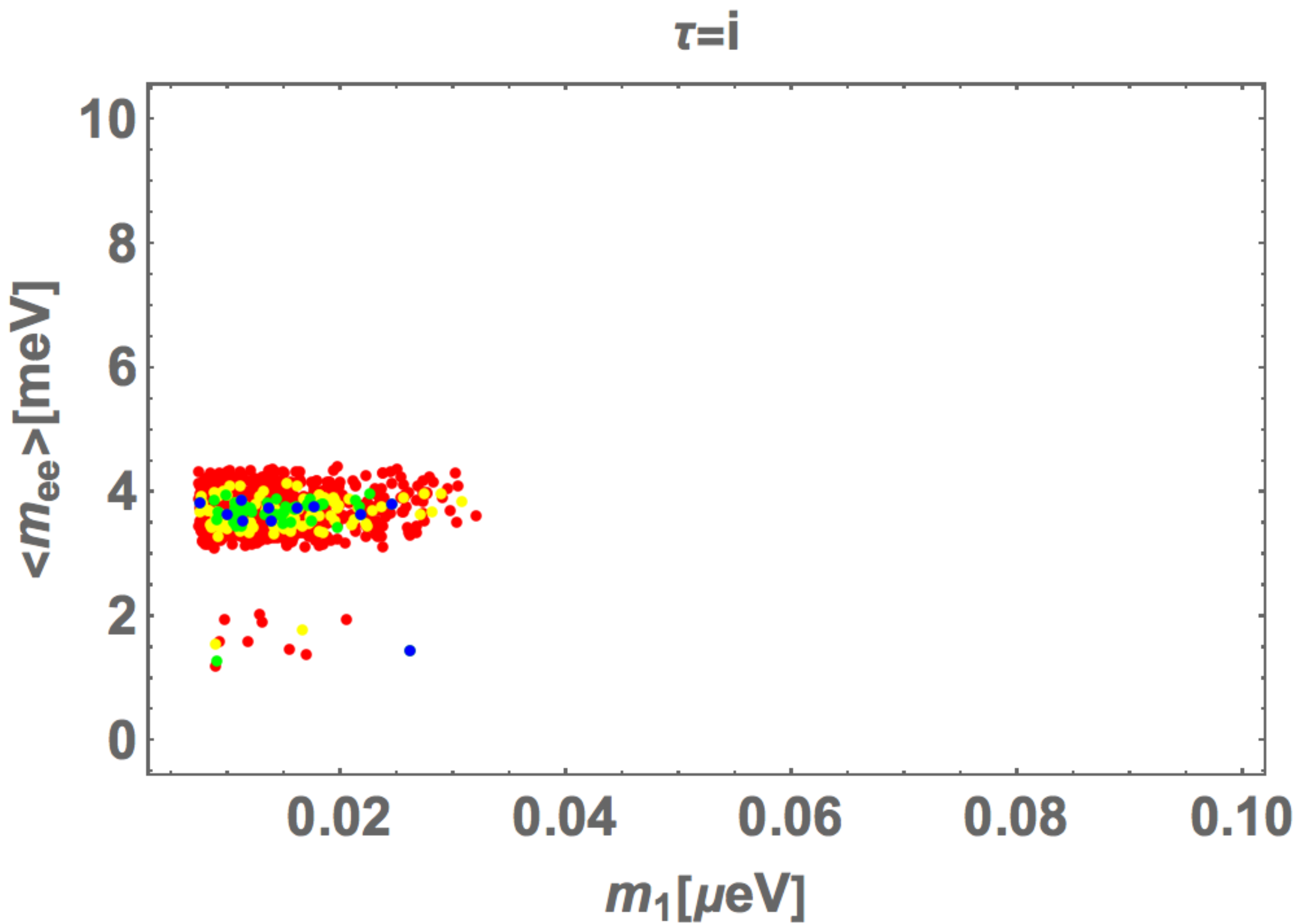}\\
  \includegraphics[width=80mm]{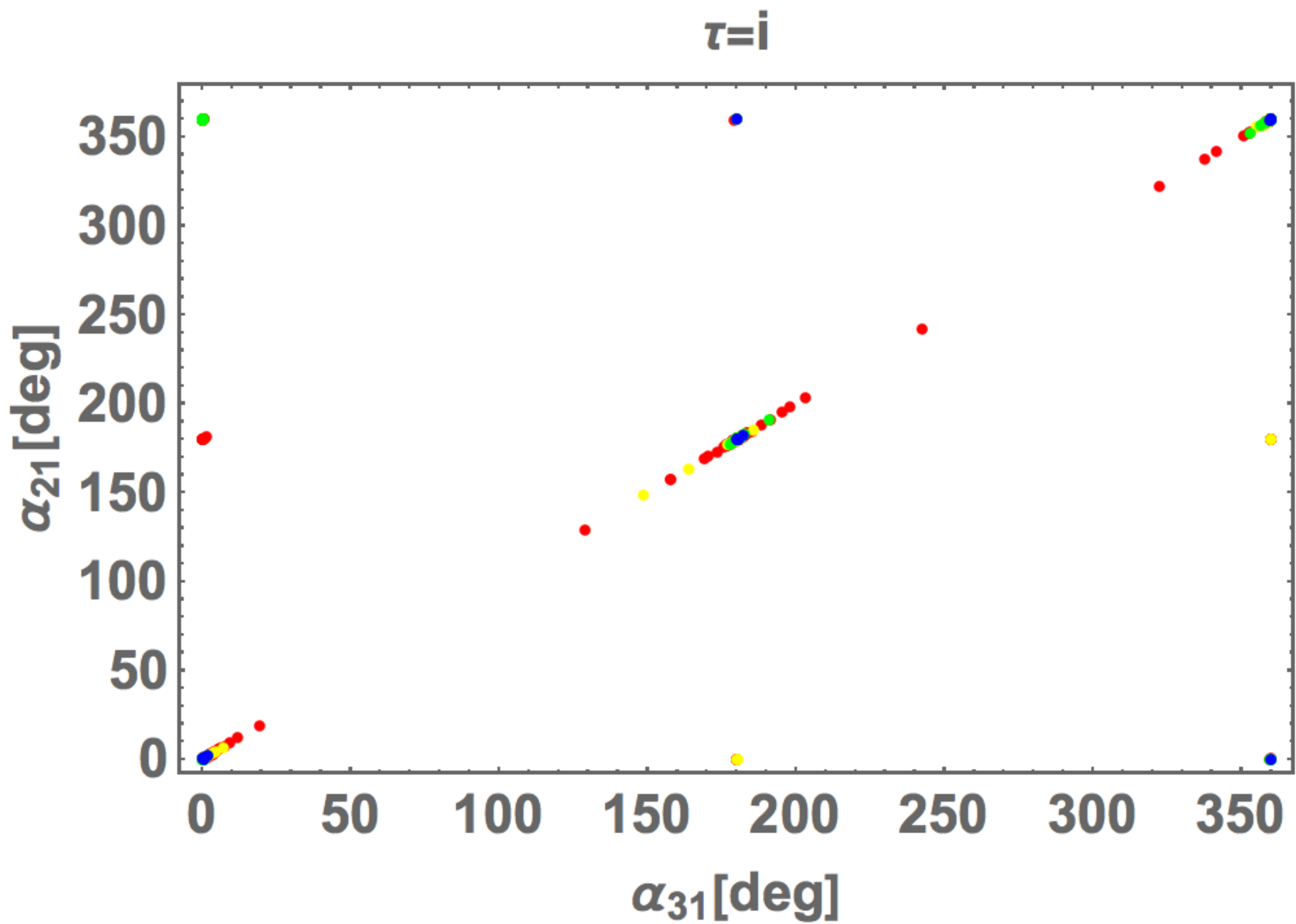}  
 \includegraphics[width=80mm]{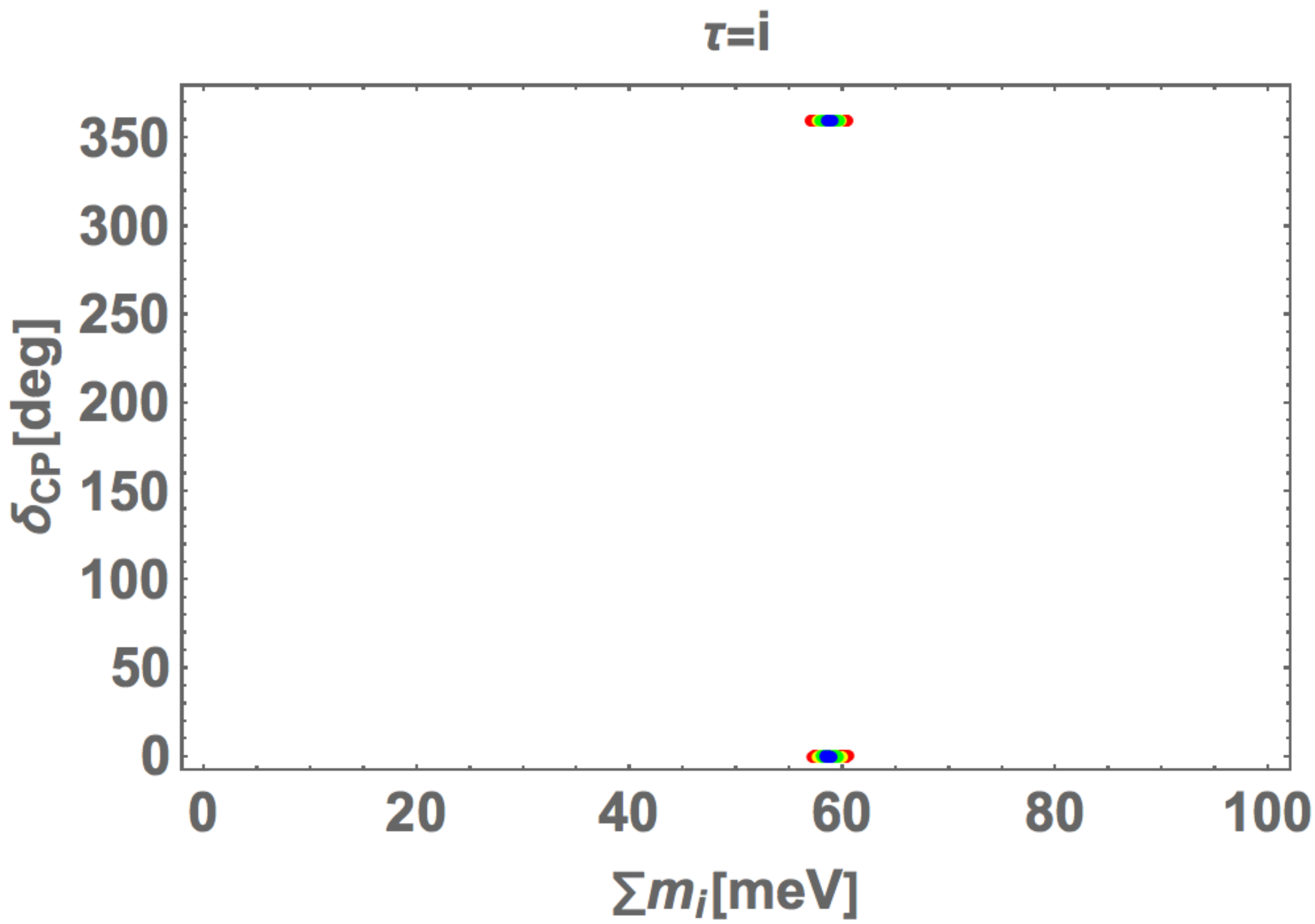}
\caption{
{By focusing on nearby $\tau=i$, we show an allowed region of $\tau$ in the top left panel, $\langle m_{ee}\rangle$ in terms of the lightest neutrino mass $m_1$ in the top right one , Majorana phases $\alpha_{21},\alpha_{31}$ in the bottom left one, and Dirac CP phase $\delta_{\rm CP}$ versus sum of neutrino masses $\sum m_i$ in the bottom right one, respectively.}}
\label{fig-i}
\end{center}\end{figure}

In case where the region is at nearby $\tau=i$, we show four plots in Fig.~\ref{fig-i}.
Each of color represents $blue\le1\sigma$, $1\sigma<green\le 2\sigma$, $2\sigma<yellow\le 3\sigma$, $3\sigma<red\le5\sigma$.  
The top left figure shows allowed region of $\tau$, the top right one $\langle m_{ee}\rangle$ in terms of the lightest neutrino mass $m_1$,
the bottom left one Majorana phases $\alpha_{21},\alpha_{31}$, and the bottom right one Dirac CP phase $\delta_{\rm CP}$ versus sum of neutrino masses $\sum m_i$. 
These figures suggest $1{\rm meV}\lesssim\langle m_{ee}\rangle \lesssim 4.5{\rm meV}$, $ m_1 \lesssim0.033{\rm \mu eV}$, 
$57{\rm meV}\lesssim \sum m_i\lesssim 61{\rm meV}$, $\delta_{\rm CP}\simeq 0^\circ$.
Allowed regions of Majorana phases tend to be localized at nearby $\alpha_{21}=\alpha_{31}$ and  $\alpha_{21}=\alpha_{31}=0$.

Also, we show a benchmark point in case of {nearby} $\tau=i$ in Table~\ref{bp-tab_i} that provides minimum $\sqrt{\Delta\chi^2}$ in our numerical analysis. 

\begin{table}[h]
	\centering
	\begin{tabular}{|c|c|c|} \hline 
			\rule[14pt]{0pt}{0pt}
 		&  NH  \\  \hline
			\rule[14pt]{0pt}{0pt}
		$\tau$ & $-0.00199671 + 1.0136 i$       \\ \hline
		\rule[14pt]{0pt}{0pt}
%
		$[a_\eta, b_\eta,c_\eta,d_\eta,e_\eta,f_\eta]$ & $[0.0343, -0.0396, 0.00351, -0.00486, -0.033, 0.0000357]$   \\ \hline
		\rule[14pt]{0pt}{0pt}
		$[\alpha_{NS},\beta_{NS}] $ & $[-0.111935, 0.00113346]$     \\ \hline
		\rule[14pt]{0pt}{0pt}
				$[s_{\theta_R},s_{\theta_I},s_{\tilde\theta}]$ & $[-0.0000658, 0.000751, -0.535]$     \\ \hline
		\rule[14pt]{0pt}{0pt}
				$[A_{\alpha_{NS}},A_{\beta_{NS}}]/{\rm GeV}$ & $[6.12\times10^{5}, 1.49\times10^{4}]$     \\ \hline
		\rule[14pt]{0pt}{0pt}
		$[M_0,m_{\tilde S},\mu_{\tilde SB},\mu_\chi]/{\rm GeV}$ & $[2.16\times10^{4}, 3.41\times10^{5}, 998, 0.000250]$     \\ \hline
		\rule[14pt]{0pt}{0pt}
		$[m_{1_R}, m_{1_I},m_{2_R}, m_{2_I},m_{\tilde \xi_1},m_{\tilde \xi_2}]/{\rm GeV}$ & $[1.54\times10^{4}, 8.75\times10^{5}, 242.552, 262.181, 2049.28, 5329.91]$    \\ \hline
		\rule[14pt]{0pt}{0pt}
		$\Delta m^2_{\rm atm}$  &  $2.51\times10^{-3} {\rm eV}^2$   \\ \hline
		\rule[14pt]{0pt}{0pt}
		$\Delta m^2_{\rm sol}$  &  $7.36\times10^{-5} {\rm eV}^2$        \\ \hline
		\rule[14pt]{0pt}{0pt}
		$\sin\theta_{12}$ & $ 0.540$   \\ \hline
		\rule[14pt]{0pt}{0pt}
		$\sin\theta_{23}$ &  $ 0.762$   \\ \hline
		\rule[14pt]{0pt}{0pt}
		$\sin\theta_{13}$ &  $ 0.151$   \\ \hline
		\rule[14pt]{0pt}{0pt}
		$[\delta_{\rm CP}^\ell,\ \alpha_{21},\,\alpha_{31}]$ &  $[0.0242851^\circ,\, 180.081^\circ,\, 180.131^\circ]$   \\ \hline
		\rule[14pt]{0pt}{0pt}
		$\sum m_i$ &  $58.7$\,meV      \\ \hline
		\rule[14pt]{0pt}{0pt}
		$\langle m_{ee} \rangle$ &  $3.59$\,meV      \\ \hline
		\rule[14pt]{0pt}{0pt}
		$\sqrt{\Delta\chi^2}$ &  $1.59$     \\ \hline
		\hline
	\end{tabular}
	\caption{Numerical {benchmark point (BP)} of our input parameters and observables at nearby the fixed point $\tau= i$ in NH. Here, this BP is taken such that $\sqrt{\Delta \chi^2}$ is minimum.}
	\label{bp-tab_i}
\end{table}

\clearpage

\subsection{$\tau=\omega$} 
\begin{figure}[tb!]\begin{center}
\includegraphics[width=80mm]{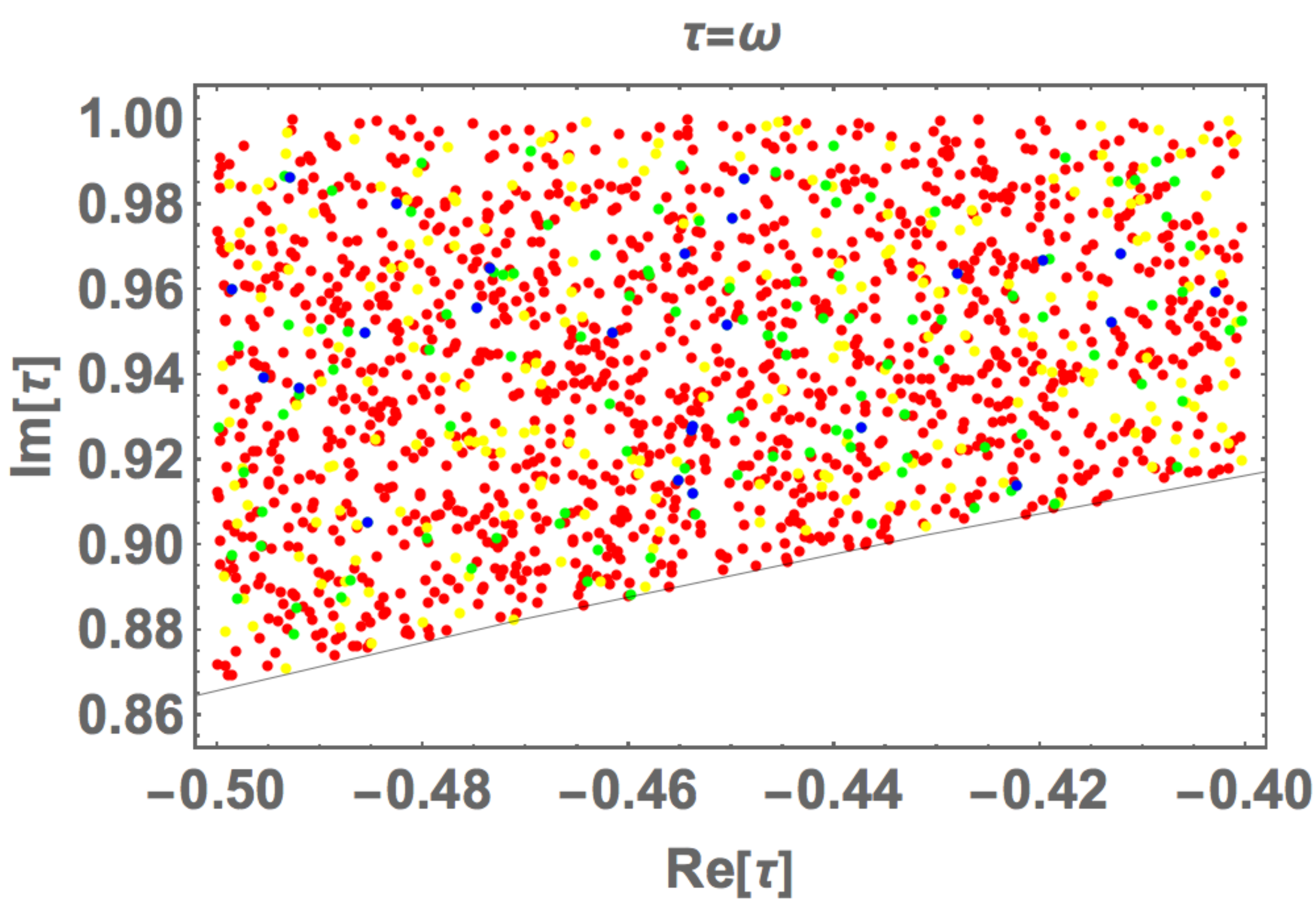} 
 \includegraphics[width=80mm]{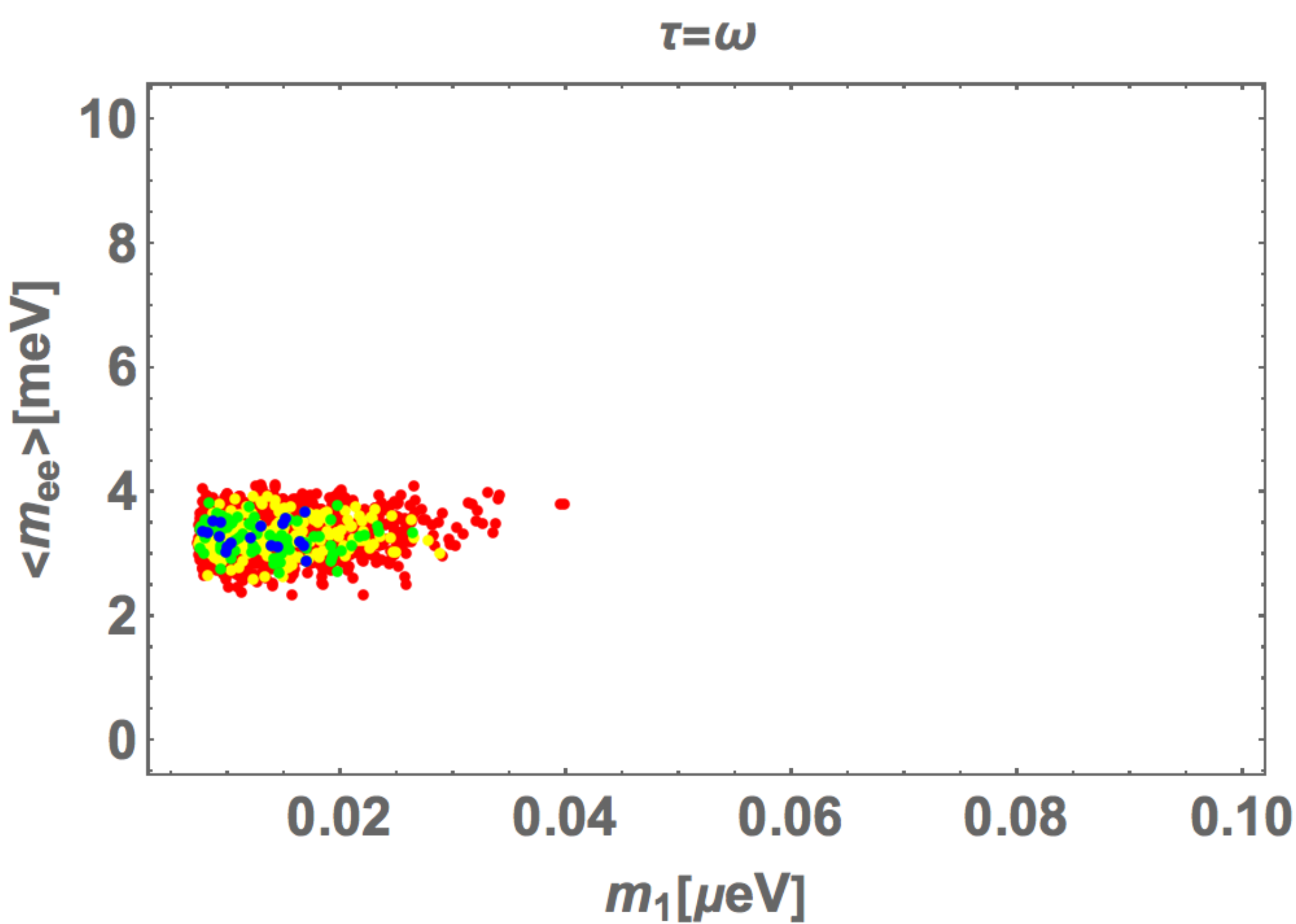}\\
  \includegraphics[width=80mm]{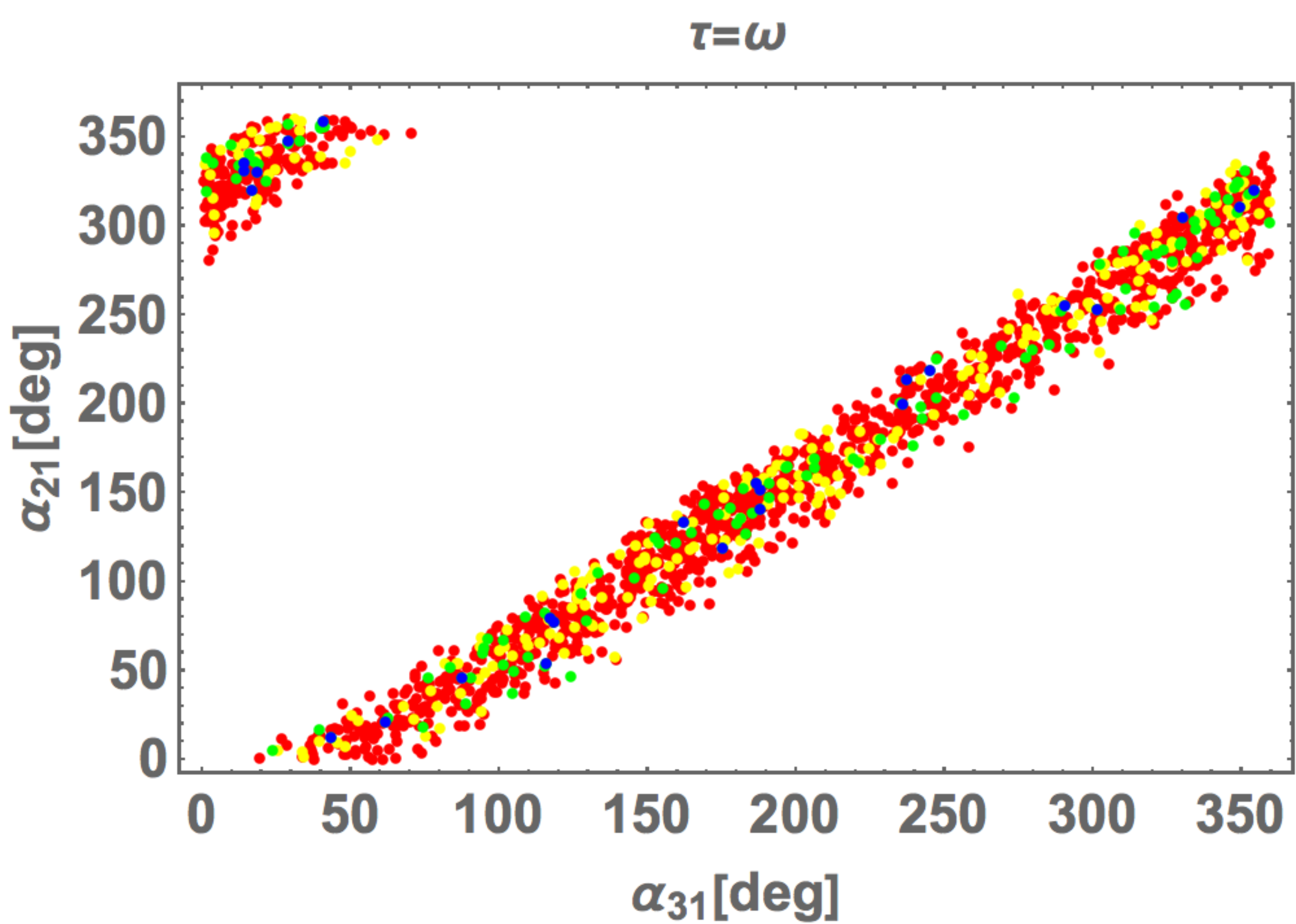}  
 \includegraphics[width=80mm]{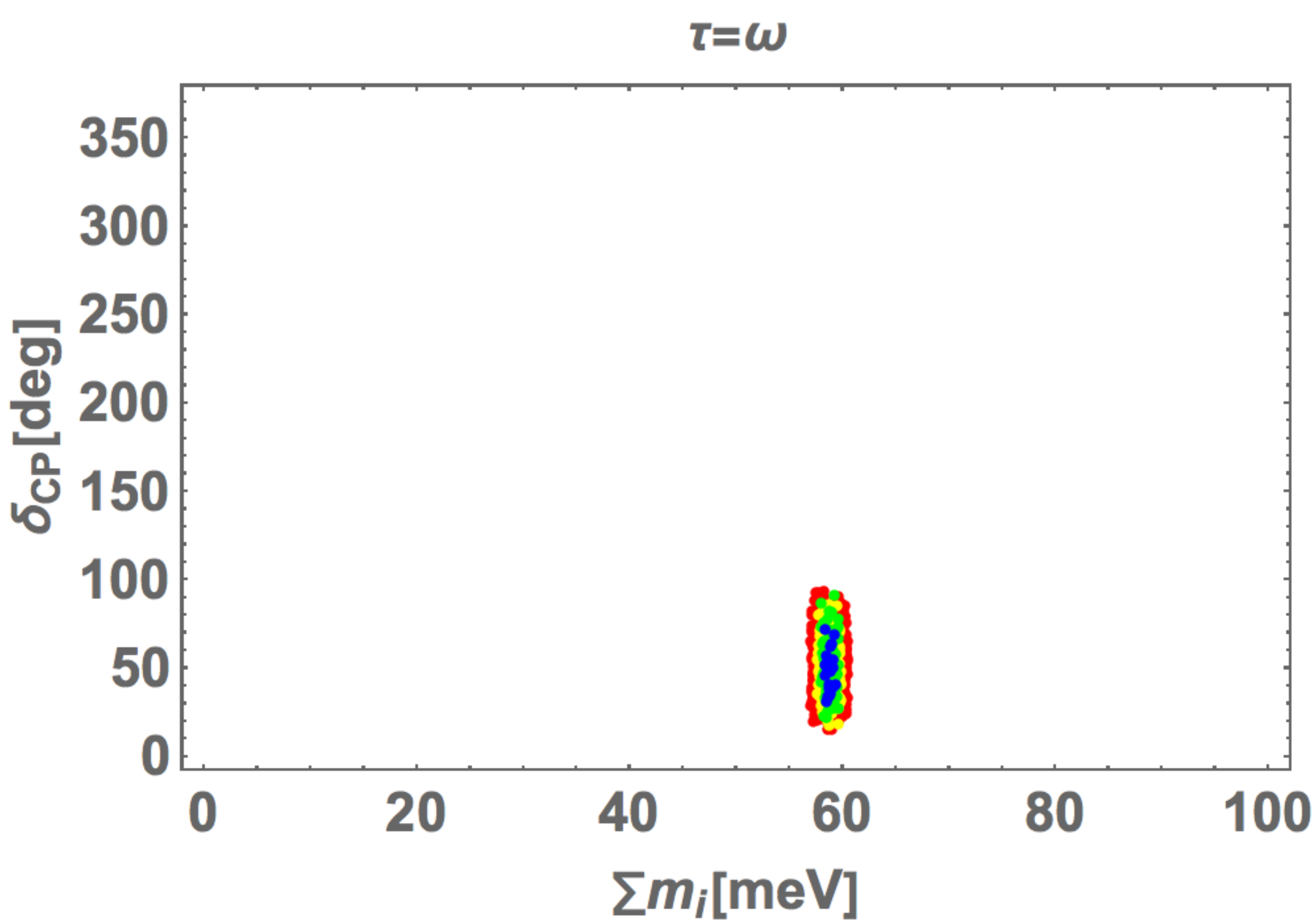}
\caption{{We analyze nearby the fixed point $\tau=\omega$, where the legends and the colors are the same as the case of nearby $\tau=i$.}}   
\label{fig-omega}
\end{center}\end{figure}

In case where the region is at nearby $\tau=\omega$, we show four plots in Fig.~\ref{fig-omega}.
The legends and the colors are the same as the case of {nearby} $\tau=i$.
These figures suggest $2.5{\rm meV}\lesssim\langle m_{ee}\rangle \lesssim 4{\rm meV}$, $ m_1 \lesssim0.04{\rm \mu eV}$, $56{\rm meV}\lesssim \sum m_i\lesssim 61{\rm meV}$, $10^\circ\lesssim \delta_{\rm CP}\lesssim 100^\circ$.
Allowed regions of Majorana phases tend to be localized at nearby $\alpha_{21}=\alpha_{31}/2$, $\alpha_{21}=\alpha_{31}\simeq0^\circ$.

Also, we show a benchmark point in case of {nearby} $\tau=\omega$ in Table~\ref{bp-tab_omega} that provides minimum $\sqrt{\chi^2}$ in our numerical analysis. 

\begin{table}[h]
	\centering
	\begin{tabular}{|c|c|c|} \hline 
			\rule[14pt]{0pt}{0pt}
 		&  NH  \\  \hline
			\rule[14pt]{0pt}{0pt}
		$\tau$ & $-0.486 + 0.950 i$       \\ \hline
		\rule[14pt]{0pt}{0pt}
%
		$[a_\eta, b_\eta,c_\eta,d_\eta,e_\eta,f_\eta]\times 10^6$ & $[-0.0936343, 935.953, 723.884, 216.329, 1023.81, -266.283]$   \\ \hline
		\rule[14pt]{0pt}{0pt}
		$[\alpha_{NS},\beta_{NS}]\times 10^9$ & $[-10208.4, 247.548]$     \\ \hline
		\rule[14pt]{0pt}{0pt}
				$[s_{\theta_R},s_{\theta_I},s_{\tilde\theta}]$ & $[-0.0000420939, 0.000104172, -0.000110707]$     \\ \hline
		\rule[14pt]{0pt}{0pt}
				$[A_{\alpha_{NS}},A_{\beta_{NS}}]/{\rm GeV}$ & $[-270895, 10427]$     \\ \hline
		\rule[14pt]{0pt}{0pt}
		$[M_0,m_{\tilde S},\mu_{\tilde SB},\mu_\chi]/{\rm GeV}$ & $[1.02274 {\times} 10^7, 160929., 589426,0.177798]$     \\ \hline
		\rule[14pt]{0pt}{0pt}
		$[m_{1_R}, m_{1_I},m_{2_R}, m_{2_I},m_{\tilde \xi_1},m_{\tilde \xi_2}]/{\rm GeV}$ & $[591.209, 3.91496\times10^6, 97591.1, 104330, 1429.83, 403217]$    \\ \hline
		\rule[14pt]{0pt}{0pt}
		$\Delta m^2_{\rm atm}$  &  $2.53\times10^{-3} {\rm eV}^2$   \\ \hline
		\rule[14pt]{0pt}{0pt}
		$\Delta m^2_{\rm sol}$  &  $7.27\times10^{-5} {\rm eV}^2$        \\ \hline
		\rule[14pt]{0pt}{0pt}
		$\sin\theta_{12}$ & $ 0.551$   \\ \hline
		\rule[14pt]{0pt}{0pt}
		$\sin\theta_{23}$ &  $ 0.750$   \\ \hline
		\rule[14pt]{0pt}{0pt}
		$\sin\theta_{13}$ &  $ 0.152$   \\ \hline
		\rule[14pt]{0pt}{0pt}
		$[\delta_{\rm CP}^\ell,\ \alpha_{21},\,\alpha_{31}]$ &  $[62.337^\circ,\, 252.618^\circ,\, 301.47^\circ]$   \\ \hline
		\rule[14pt]{0pt}{0pt}
		$\sum m_i$ &  $58.8$\,meV      \\ \hline
		\rule[14pt]{0pt}{0pt}
		$\langle m_{ee} \rangle$ &  $3.03$\,meV      \\ \hline
		\rule[14pt]{0pt}{0pt}
		$\sqrt{\Delta\chi^2}$ &  $1.45$     \\ \hline
		\hline
	\end{tabular}
	\caption{Numerical {BP} of our input parameters and observables at nearby the fixed point $\tau=\omega$ in NH. Here, this BP is taken such that $\sqrt{\Delta \chi^2}$ is minimum.}
	\label{bp-tab_omega}
\end{table}

\clearpage

\subsection{$\tau=i\times \infty$} 
\begin{figure}[tb!]\begin{center}
\includegraphics[width=80mm]{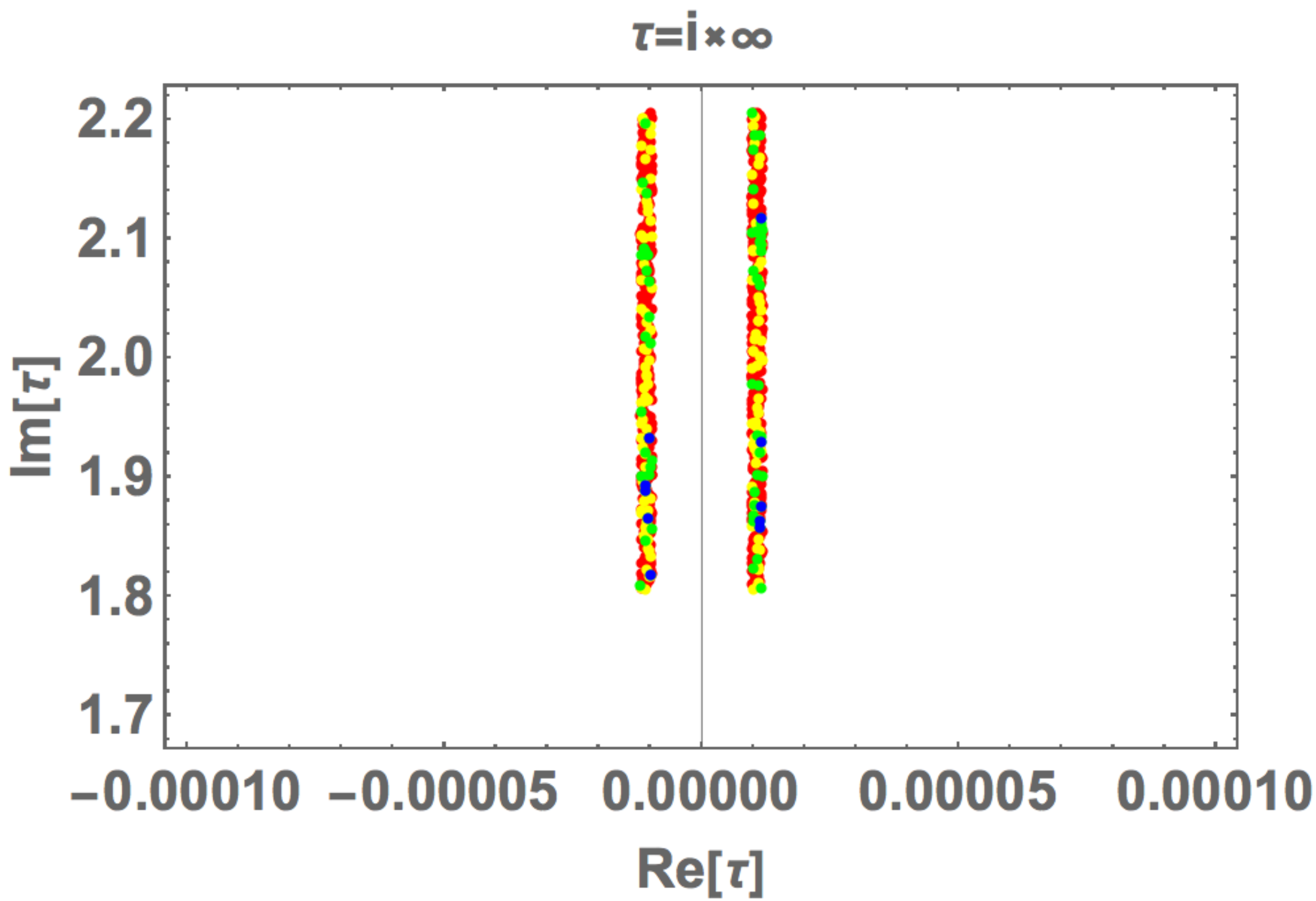} 
 \includegraphics[width=80mm]{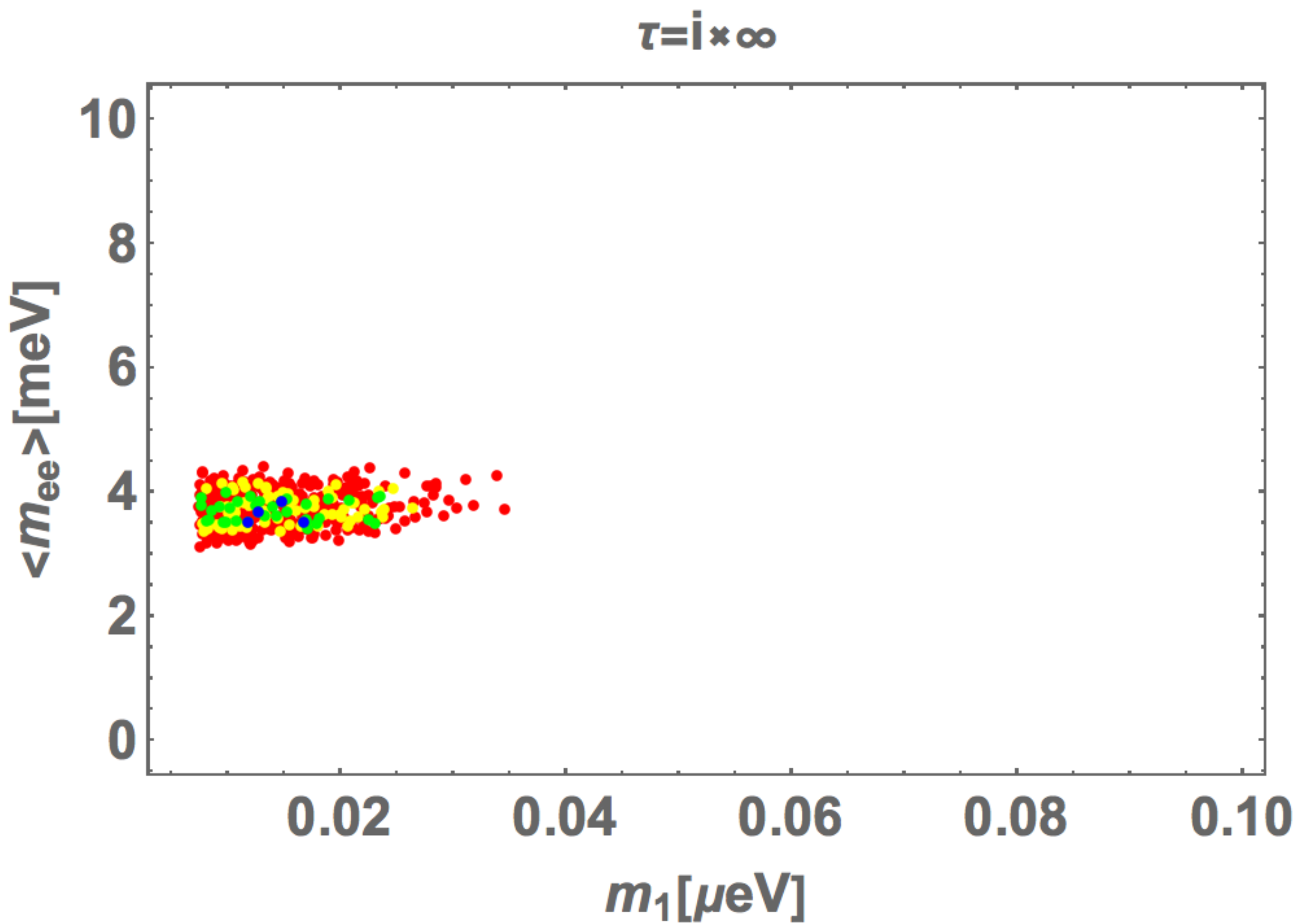}\\
  \includegraphics[width=80mm]{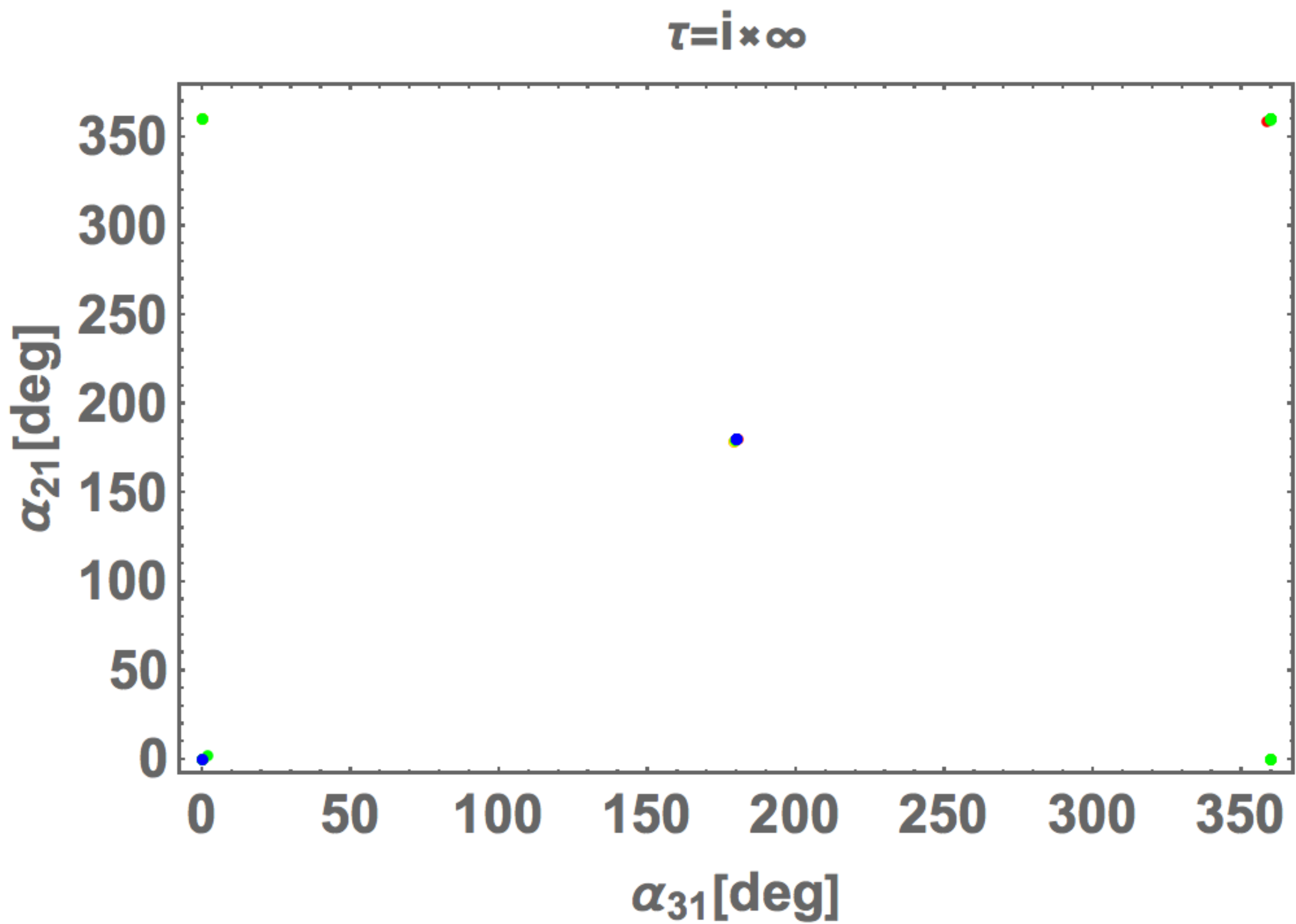}  
 \includegraphics[width=80mm]{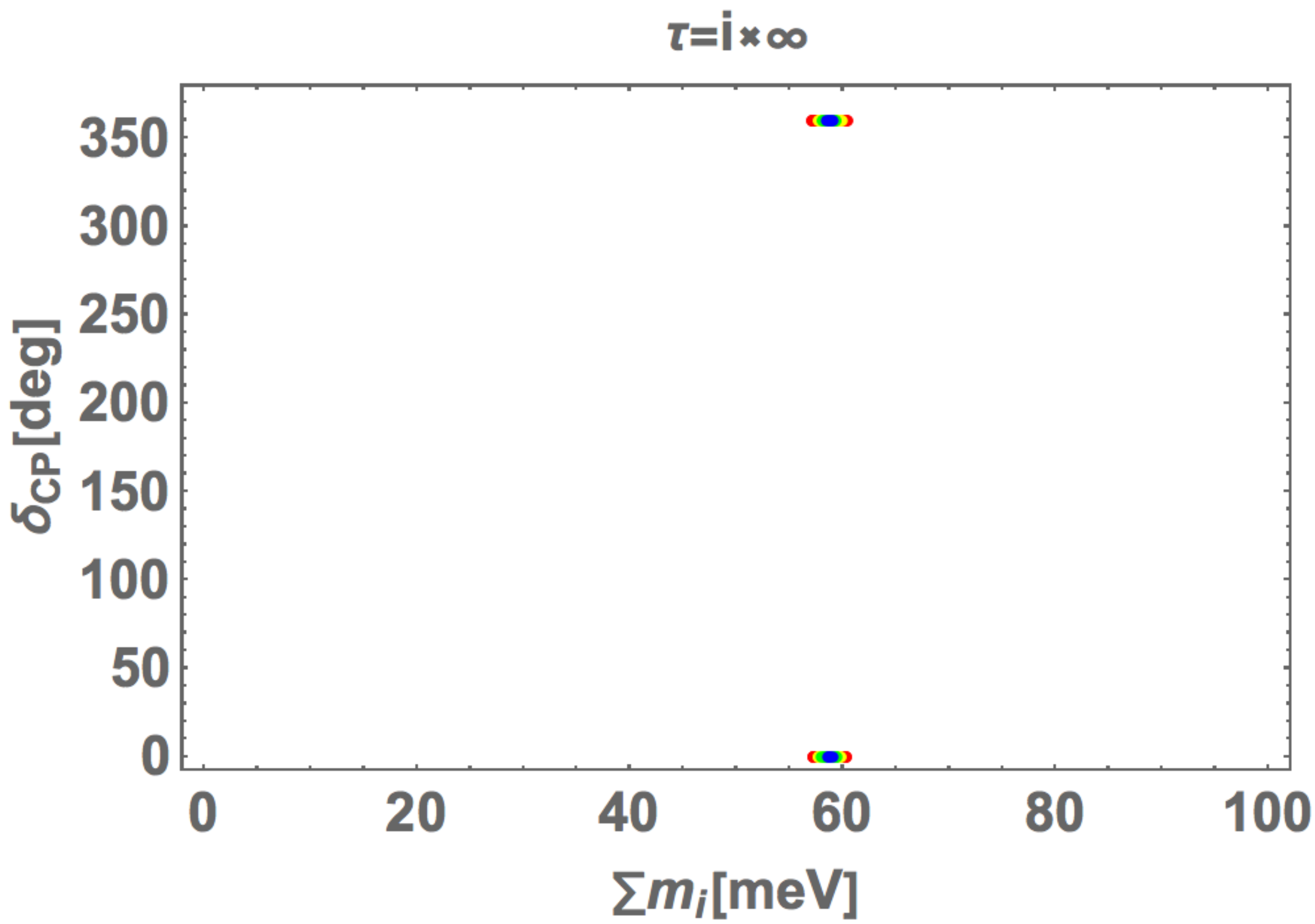}
\caption{{We analyze nearby the fixed point $\tau=i\times \infty$, where the legends and the colors are the same as the case of nearby $\tau=i$.}}   
\label{fig-infty}
\end{center}\end{figure}

In case where the region is at nearby $\tau=i\times \infty$, we show four plots in Fig.~\ref{fig-infty}.
The legends and the colors are the same as the case of {nearby} $\tau=i$.
These figures suggest $3.0{\rm meV}\lesssim\langle m_{ee}\rangle \lesssim 4.5{\rm meV}$, $ m_1 \lesssim0.035{\rm \mu eV}$, 
$57{\rm meV}\lesssim \sum m_i\lesssim 61{\rm meV}$, $\delta_{\rm CP}\simeq 0^\circ$. 
Majorana phases are localized at nearby $\alpha_{21}=\alpha_{31}=[0^\circ,\   180^\circ]$.

Also, we show a benchmark point in case of {nearby} $\tau=i\times \infty$ in Table~\ref{bp-tab_infty} that provides minimum $\sqrt{\chi^2}$ in our numerical analysis. 

\begin{table}[h]
	\centering
	\begin{tabular}{|c|c|c|} \hline 
			\rule[14pt]{0pt}{0pt}
 		&  NH  \\  \hline
			\rule[14pt]{0pt}{0pt}
		$\tau$ & $-9.91\times10^{-6} + 1.82 i$       \\ \hline
		\rule[14pt]{0pt}{0pt}
%
		$[a_\eta, b_\eta,c_\eta,d_\eta,e_\eta,f_\eta] $ & $[-0.000569, -0.0000728, -0.002268, -0.00161, 0.00134, \
-0.000562]$   \\ \hline
		\rule[14pt]{0pt}{0pt}
		$[\alpha_{NS},\beta_{NS}] $ & $[0.00228, 0.000373]$     \\ \hline
		\rule[14pt]{0pt}{0pt}
				$[s_{\theta_R},s_{\theta_I},s_{\tilde\theta}]$ & $[0.00232, 0.0000629, 0.000992]$     \\ \hline
		\rule[14pt]{0pt}{0pt}
				$[A_{\alpha_{NS}},A_{\beta_{NS}}]/{\rm GeV}$ & $[1.43\times10^5, 1450]$     \\ \hline
		\rule[14pt]{0pt}{0pt}
		$[M_0,m_{\tilde S},\mu_{\tilde SB},\mu_\chi]/{\rm GeV}$ & $[4.87\times10^6, 2050, 2120, 0.00133]$     \\ \hline
		\rule[14pt]{0pt}{0pt}
		$[m_{1_R}, m_{1_I},m_{2_R}, m_{2_I},m_{\tilde \xi_1},m_{\tilde \xi_2}]/{\rm GeV}$ & $[1.09\times10^6, 1.19\times10^5, 8.86\times10^4, 9.53\times10^4, 1.14\times10^4, 5.08\times10^4]$    \\ \hline
		\rule[14pt]{0pt}{0pt}
		$\Delta m^2_{\rm atm}$  &  $2.52\times10^{-3} {\rm eV}^2$   \\ \hline
		\rule[14pt]{0pt}{0pt}
		$\Delta m^2_{\rm sol}$  &  $7.37\times10^{-5} {\rm eV}^2$        \\ \hline
		\rule[14pt]{0pt}{0pt}
		$\sin\theta_{12}$ & $ 0.537$   \\ \hline
		\rule[14pt]{0pt}{0pt}
		$\sin\theta_{23}$ &  $ 0.764$   \\ \hline
		\rule[14pt]{0pt}{0pt}
		$\sin\theta_{13}$ &  $ 0.147$   \\ \hline
		\rule[14pt]{0pt}{0pt}
		$[\delta_{\rm CP}^\ell,\ \alpha_{21},\,\alpha_{31}]$ &  $[0^\circ,\, 180^\circ,\, 180^\circ]$   \\ \hline
		\rule[14pt]{0pt}{0pt}
		$\sum m_i$ &  $58.7$\,meV      \\ \hline
		\rule[14pt]{0pt}{0pt}
		$\langle m_{ee} \rangle$ &  $3.38$\,meV      \\ \hline
		\rule[14pt]{0pt}{0pt}
		$\sqrt{\Delta\chi^2}$ &  $1.68$     \\ \hline
		\hline
	\end{tabular}
	\caption{Numerical {BP} of our input parameters and observables at nearby the fixed point $\tau=i\times\infty$ in NH. Here, this BP is taken such that $\sqrt{\Delta \chi^2}$ is minimum.}
	\label{bp-tab_infty}
\end{table}

\section{Conclusion and discussion}
\label{sec:conclusion}

We have studied a modular $A_4$ invariant two-loop neutrino mass model in a SUSY framework,
in which we have focused on regions at nearby three fixed points of $\tau=i,\omega,i\times \infty$.
{These points with residual symmetries are motivated by flux compactifications of the string theory.}

Thanks to contributions of SUSY partners $\tilde N^c, \tilde\chi,\tilde\eta_1$ to the neutrino mass matrix, we have successfully obtained several predictions in NH such as phases and neutrino masses for each of fixed points through our global $\Delta \chi^2$ analysis.
Moreover, the non-SUSY contributions to the neutrino masses are negligibly small at some points, since the mass of $N^c$ is minuscule
in addition to rather small Yukawa couplings $10^{-2}\sim10^{-3}$. Here, $10^{-3}$ comes from the conservative upper limit to satisfy the lepton flavor violations such as $\mu\to e\gamma$.
On the other hand, the mass of $\tilde N^c$ can be larger than the mass of $N^c$, which could reach at the order TeV, due to soft {SUSY-breaking} terms such as $A_{\alpha_{NS}}, A_{\beta_{NS}}, m_{\tilde S}, \mu_{\tilde SB}$.

Before closing our paper, we will briefly mention dark matter (DM) candidates.
In our model, the mass of $N^c$ (as well as its superpartner) is induced at one-loop level.~\footnote{To make our discussion simple, we focus on non-SUSY particles only.}
It implies that the mass of DM candidate is naturally smaller than the other particles.
But the problem would arise from the thermal averaged cross section to satisfy the observed relic density of DM since valid interactions come from Yukawa couplings only, and their order is $10^{-2}$ at most.
In fact, order one Yukawa couplings are needed in order to explain the relic density.
Thus, we need to rely on bosonic DM candidate; i.e., $\chi$ or neutral component of $\eta_{1,2}$~\cite{Nagao:2021rio}.
Since $\chi$ and $\eta_1$ directly interact with $N^c$ through Yukawa couplings and $\eta_2$ mixes with $\chi$ and $\eta_1$, one component DM is favored when the bosonic DM mass is lighter than the mass of $N^c$.
For simplicity, let us consider $\eta\equiv \eta_{1,2}$ dominant DM or $\chi$ dominant DM, neglecting the mixing among them.
In case where $\eta$ is the DM candidate, the main interaction comes from kinetic terms and it is known that there exist some solutions to satisfy the relic density. Here we cannot rely on the Yukawa interactions due to the similar reason as the case of fermionic DM.
According to the systematic analysis by ref.~\cite{Hambye:2009pw}, it tells us the dark matter mass is at around 534 GeV when the mass is larger than the mass of $W/Z$ mass.
In the lighter region, one also finds the dark matter mass is at around the half of Higgs mass; 63 GeV.
In order to avoid the constraint of direct detection searches, we need mass difference between the real and imaginary part of $\eta$ that is more than 100 keV.  It implies that we need a little mixings among neutral bosons.
In case where $\chi$ is the DM candidate, the main interaction comes from Higgs potential, and any mass range could be possible through these interactions~\cite{Kanemura:2010sh}.

\section*{Acknowledgments}
This work was supported by JSPS KAKENHI Grant Numbers JP19J00664 (Hajime O.) and JP20K14477 (Hajime O.). 
The work of Hiroshi O. is supported by the Junior Research Group (JRG) Program at the Asia-Pacific Center for Theoretical
Physics (APCTP) through the Science and Technology Promotion Fund and Lottery Fund of the Korean Government and was supported by the Korean Local Governments-Gyeongsangbuk-do Province and Pohang City. 
Hiroshi O. is sincerely grateful for all the KIAS members.

{

 \appendix

\section{Formulas in modular $A_4$ framework}

In this appendix, we summarize some formulas in the framework of $A_4$ modular symmetry belonging to 
the $SL(2,\mathbb{Z})$ symmetry. 
The modulus $\tau$ transforms as
\begin{align}
& \tau \longrightarrow \gamma\tau= \frac{a\tau + b}{c \tau + d},
\end{align}
with $\{a,b,c,d\} \in \mathbb{Z}$ satisfying $ad-bc=1$ and ${\rm Im} [\tau]>0$.
The transformation of modular forms $f(\tau)$ are given by
\begin{align}
& f(\gamma\tau)= (c\tau+d)^k f(\tau)~, ~~ \gamma \in \Gamma(N)~ ,
\end{align}
where $f(\tau)$ denotes holomorphic functions of $\tau$ with the modular weight $k$.

In a similar way, the modular transformation of a matter chiral superfield $\phi^{(I)}$ with the modular weight $-k_I$ 
is given by 
\begin{equation}
\phi^{(I)} \to (c\tau+d)^{-k_I}\rho^{(I)}(\gamma)\phi^{(I)},
\end{equation}
where $\rho^{(I)}(\gamma)$ stands for an unitary matrix corresponding to $A_4$ transformation. 
Note that the superpotential is invariant when the sum of modular weight from fields and modular form is zero. 
It restricts a form of the superpotential as shown in Eq. (\ref{eq:sp-lep}).

Modular forms are constructed on the basis of weight 2 modular form, $ Y^{(2)}_3=  (y_{1},y_{2},y_{3})$, transforming
as a triplet of $A_4$. 
Their explicit forms are written by the Dedekind eta-function $\eta(\tau)$ and its derivative with respect to $\tau$ \cite{Feruglio:2017spp}:
\begin{eqnarray} 
\label{eq:Y-A4}
y_{1}(\tau) &=& \frac{i}{2\pi}\left( \frac{\eta'(\tau/3)}{\eta(\tau/3)}  +\frac{\eta'((\tau +1)/3)}{\eta((\tau+1)/3)}  
+\frac{\eta'((\tau +2)/3)}{\eta((\tau+2)/3)} - \frac{27\eta'(3\tau)}{\eta(3\tau)}  \right), \nonumber \\
y_{2}(\tau) &=& \frac{-i}{\pi}\left( \frac{\eta'(\tau/3)}{\eta(\tau/3)}  +\omega^2\frac{\eta'((\tau +1)/3)}{\eta((\tau+1)/3)}  
+\omega \frac{\eta'((\tau +2)/3)}{\eta((\tau+2)/3)}  \right) , \label{eq:Yi} \\ 
y_{3}(\tau) &=& \frac{-i}{\pi}\left( \frac{\eta'(\tau/3)}{\eta(\tau/3)}  +\omega\frac{\eta'((\tau +1)/3)}{\eta((\tau+1)/3)}  
+\omega^2 \frac{\eta'((\tau +2)/3)}{\eta((\tau+2)/3)}  \right)\,, \nonumber \\
 \eta(\tau) &=& q^{1/24}\Pi_{n=1}^\infty (1-q^n), \quad q=e^{2\pi i \tau}, \quad \omega=e^{2\pi i /3}.
\nonumber
\end{eqnarray}
%
Modular forms of higher weight can be obtained from tensor products of $Y^{(2)}_3$. 
We enumerate some $A_4$ singlet modular forms used in our analysis:
\begin{align}
Y_1^{(4)} &= y_1^2 + 2 y_1 y_3, \quad Y^{(4)}_{1'} = y^2_3+2 y_1 y_2,
\quad Y^{(6)}_{\bf 1}= y_1^2 + y_2^2 + y_3^2 - 3 y_1 y_2 y_3, \nonumber\\
Y_1^{(8)} &=(y_1^2+ 2 y_2 y_3)^2, \quad Y^{(8)}_{\bf 1'}=(y_1^2+ 2 y_2 y_3)(y_3^2+ 2 y_1 y_2), 
\quad Y^{(8)}_{\bf 1''}=(y_3^2+ 2 y_1 y_2)^2, 
\end{align}
where the number in superscript denotes the modular weight.
The $A_4$ triplet of the modular weight 4 is given by $Y^{(4)}_3=(y^2_1-2y_2y_3, y^2_3-2y_1y_2, y^2_2-2y_1y_3)$.


}


\begin{thebibliography}{99}

\bibitem{Ma:2006km} 
  E.~Ma,
  Phys.\ Rev.\ D {\bf 73}, 077301 (2006)
  [hep-ph/0601225].



\bibitem{Feruglio:2017spp}
F.~Feruglio,
[arXiv:1706.08749 [hep-ph]].

\bibitem{deAdelhartToorop:2011re}
R.~de Adelhart Toorop, F.~Feruglio and C.~Hagedorn,
Nucl. Phys. B \textbf{858}, 437-467 (2012)
[arXiv:1112.1340 [hep-ph]].

\bibitem{Criado:2018thu}
J.~C.~Criado and F.~Feruglio,
SciPost Phys. \textbf{5}, no.5, 042 (2018)
[arXiv:1807.01125 [hep-ph]].

\bibitem{Kobayashi:2018scp}
T.~Kobayashi, N.~Omoto, Y.~Shimizu, K.~Takagi, M.~Tanimoto and T.~H.~Tatsuishi,
JHEP \textbf{11}, 196 (2018)
[arXiv:1808.03012 [hep-ph]].

\bibitem{Okada:2018yrn}
H.~Okada and M.~Tanimoto,
Phys. Lett. B \textbf{791}, 54-61 (2019)
[arXiv:1812.09677 [hep-ph]].

\bibitem{Kobayashi:2021ajl}
T.~Kobayashi, H.~Okada and Y.~Orikasa,
[arXiv:2111.05674 [hep-ph]].

\bibitem{Nomura:2019jxj}
T.~Nomura and H.~Okada,
Phys. Lett. B \textbf{797}, 134799 (2019)
[arXiv:1904.03937 [hep-ph]].

\bibitem{Okada:2019uoy}
H.~Okada and M.~Tanimoto,
Eur. Phys. J. C \textbf{81}, no.1, 52 (2021)
[arXiv:1905.13421 [hep-ph]].

\bibitem{deAnda:2018ecu}
F.~J.~de Anda, S.~F.~King and E.~Perdomo,
Phys. Rev. D \textbf{101}, no.1, 015028 (2020)
[arXiv:1812.05620 [hep-ph]].

\bibitem{Novichkov:2018yse}
P.~P.~Novichkov, S.~T.~Petcov and M.~Tanimoto,
Phys. Lett. B \textbf{793}, 247-258 (2019)
[arXiv:1812.11289 [hep-ph]].

\bibitem{Nomura:2019yft}
T.~Nomura and H.~Okada,
Nucl. Phys. B \textbf{966}, 115372 (2021)
[arXiv:1906.03927 [hep-ph]].

\bibitem{Okada:2019mjf}
H.~Okada and Y.~Orikasa,
[arXiv:1907.13520 [hep-ph]].

\bibitem{Ding:2019zxk}
G.~J.~Ding, S.~F.~King and X.~G.~Liu,
JHEP \textbf{09}, 074 (2019)
[arXiv:1907.11714 [hep-ph]].

\bibitem{Nomura:2019lnr}
T.~Nomura, H.~Okada and O.~Popov,
Phys. Lett. B \textbf{803}, 135294 (2020)
[arXiv:1908.07457 [hep-ph]].

\bibitem{Kobayashi:2019xvz}
T.~Kobayashi, Y.~Shimizu, K.~Takagi, M.~Tanimoto and T.~H.~Tatsuishi,
Phys. Rev. D \textbf{100}, no.11, 115045 (2019)
[erratum: Phys. Rev. D \textbf{101}, no.3, 039904 (2020)]
[arXiv:1909.05139 [hep-ph]].

\bibitem{Asaka:2019vev}
T.~Asaka, Y.~Heo, T.~H.~Tatsuishi and T.~Yoshida,
JHEP \textbf{01}, 144 (2020)
[arXiv:1909.06520 [hep-ph]].

\bibitem{Zhang:2019ngf}
D.~Zhang,
Nucl. Phys. B \textbf{952}, 114935 (2020)
[arXiv:1910.07869 [hep-ph]].

\bibitem{Gui-JunDing:2019wap}
G.~J.~Ding, S.~F.~King, X.~G.~Liu and J.~N.~Lu,
JHEP \textbf{12}, 030 (2019)
[arXiv:1910.03460 [hep-ph]].

\bibitem{Kobayashi:2019gtp}
T.~Kobayashi, T.~Nomura and T.~Shimomura,
Phys. Rev. D \textbf{102}, no.3, 035019 (2020)
[arXiv:1912.00637 [hep-ph]].

\bibitem{Nomura:2019xsb}
T.~Nomura, H.~Okada and S.~Patra,
Nucl. Phys. B \textbf{967}, 115395 (2021)
[arXiv:1912.00379 [hep-ph]].

\bibitem{Wang:2019xbo}
X.~Wang,
Nucl. Phys. B \textbf{957}, 115105 (2020)
[arXiv:1912.13284 [hep-ph]].

\bibitem{Okada:2020dmb}
H.~Okada and Y.~Shoji,
Nucl. Phys. B \textbf{961}, 115216 (2020)
[arXiv:2003.13219 [hep-ph]].

\bibitem{Okada:2020rjb}
H.~Okada and M.~Tanimoto,
[arXiv:2005.00775 [hep-ph]].

\bibitem{Behera:2020lpd}
M.~K.~Behera, S.~Singirala, S.~Mishra and R.~Mohanta,
[arXiv:2009.01806 [hep-ph]].

\bibitem{Behera:2020sfe}
M.~K.~Behera, S.~Mishra, S.~Singirala and R.~Mohanta,
[arXiv:2007.00545 [hep-ph]].

\bibitem{Nomura:2020opk}
T.~Nomura and H.~Okada,
[arXiv:2007.04801 [hep-ph]].

\bibitem{Nomura:2020cog}
T.~Nomura and H.~Okada,
[arXiv:2007.15459 [hep-ph]].

\bibitem{Asaka:2020tmo}
T.~Asaka, Y.~Heo and T.~Yoshida,
Phys. Lett. B \textbf{811}, 135956 (2020)
[arXiv:2009.12120 [hep-ph]].

\bibitem{Okada:2020ukr}
H.~Okada and M.~Tanimoto,
Phys. Rev. D \textbf{103}, no.1, 015005 (2021)
[arXiv:2009.14242 [hep-ph]].

\bibitem{Nagao:2020snm}
K.~I.~Nagao and H.~Okada,
[arXiv:2010.03348 [hep-ph]].

\bibitem{Okada:2020brs}
H.~Okada and M.~Tanimoto,
JHEP \textbf{03}, 010 (2021)
[arXiv:2012.01688 [hep-ph]].

\bibitem{Yao:2020qyy}
C.~Y.~Yao, J.~N.~Lu and G.~J.~Ding,
JHEP \textbf{05} (2021), 102
[arXiv:2012.13390 [hep-ph]].

\bibitem{Chen:2021zty}
P.~Chen, G.~J.~Ding and S.~F.~King,
JHEP \textbf{04} (2021), 239
[arXiv:2101.12724 [hep-ph]].

\bibitem{Kashav:2021zir}
M.~Kashav and S.~Verma,
[arXiv:2103.07207 [hep-ph]].

\bibitem{Okada:2021qdf}
H.~Okada, Y.~Shimizu, M.~Tanimoto and T.~Yoshida,
[arXiv:2105.14292 [hep-ph]].

\bibitem{deMedeirosVarzielas:2021pug}
I.~de Medeiros Varzielas and J.~Louren\c{c}o,
[arXiv:2107.04042 [hep-ph]].

\bibitem{Nomura:2021yjb}
T.~Nomura, H.~Okada and Y.~Orikasa,
[arXiv:2106.12375 [hep-ph]].

\bibitem{Hutauruk:2020xtk}
P.~T.~P.~Hutauruk, D.~W.~Kang, J.~Kim and H.~Okada,
[arXiv:2012.11156 [hep-ph]].

\bibitem{Ding:2021eva}
G.~J.~Ding, S.~F.~King and J.~N.~Lu,
[arXiv:2108.09655 [hep-ph]].

\bibitem{Nagao:2021rio}
K.~I.~Nagao and H.~Okada,
[arXiv:2108.09984 [hep-ph]].


\bibitem{king}
Georgianna Charalampous, Stephen F. King, George K. Leontaris, Ye-Ling Zhou
[arXiv:2109.11379 [hep-ph]].

\bibitem{Okada:2021aoi}
H.~Okada and Y.~h.~Qi,
[arXiv:2109.13779 [hep-ph]].

\bibitem{Nomura:2021pld}
T.~Nomura, H.~Okada and Y.~h.~Qi,
[arXiv:2111.10944 [hep-ph]].

\bibitem{Kobayashi:2021pav}
T.~Kobayashi, H.~Otsuka, M.~Tanimoto and K.~Yamamoto,
[arXiv:2112.00493 [hep-ph]].

\bibitem{Dasgupta:2021ggp}
A.~Dasgupta, T.~Nomura, H.~Okada, O.~Popov and M.~Tanimoto,
[arXiv:2111.06898 [hep-ph]].

\bibitem{Liu:2021gwa}
X.~G.~Liu and G.~J.~Ding,
[arXiv:2112.14761 [hep-ph]].



\bibitem{Nomura:2022hxs}
T.~Nomura and H.~Okada,
[arXiv:2201.10244 [hep-ph]].


\bibitem{Kobayashi:2018vbk}
T.~Kobayashi, K.~Tanaka and T.~H.~Tatsuishi,
Phys. Rev. D \textbf{98}, no.1, 016004 (2018)
[arXiv:1803.10391 [hep-ph]].

\bibitem{Kobayashi:2018wkl}
T.~Kobayashi, Y.~Shimizu, K.~Takagi, M.~Tanimoto, T.~H.~Tatsuishi and H.~Uchida,
Phys. Lett. B \textbf{794}, 114-121 (2019)
[arXiv:1812.11072 [hep-ph]].

\bibitem{Kobayashi:2019rzp}
T.~Kobayashi, Y.~Shimizu, K.~Takagi, M.~Tanimoto and T.~H.~Tatsuishi,
PTEP \textbf{2020}, no.5, 053B05 (2020)
[arXiv:1906.10341 [hep-ph]].

\bibitem{Okada:2019xqk}
H.~Okada and Y.~Orikasa,
Phys. Rev. D \textbf{100}, no.11, 115037 (2019)
[arXiv:1907.04716 [hep-ph]].

\bibitem{Mishra:2020gxg}
S.~Mishra,
[arXiv:2008.02095 [hep-ph]].

\bibitem{Du:2020ylx}
X.~Du and F.~Wang,
JHEP \textbf{02}, 221 (2021)
[arXiv:2012.01397 [hep-ph]].

\bibitem{Penedo:2018nmg}
J.~T.~Penedo and S.~T.~Petcov,
Nucl. Phys. B \textbf{939}, 292-307 (2019)
[arXiv:1806.11040 [hep-ph]].

\bibitem{Novichkov:2018ovf}
P.~P.~Novichkov, J.~T.~Penedo, S.~T.~Petcov and A.~V.~Titov,
JHEP \textbf{04}, 005 (2019)
[arXiv:1811.04933 [hep-ph]].

\bibitem{Kobayashi:2019mna}
T.~Kobayashi, Y.~Shimizu, K.~Takagi, M.~Tanimoto and T.~H.~Tatsuishi,
JHEP \textbf{02}, 097 (2020)
[arXiv:1907.09141 [hep-ph]].

\bibitem{King:2019vhv}
S.~F.~King and Y.~L.~Zhou,
Phys. Rev. D \textbf{101}, no.1, 015001 (2020)
[arXiv:1908.02770 [hep-ph]].

\bibitem{Okada:2019lzv}
H.~Okada and Y.~Orikasa,
[arXiv:1908.08409 [hep-ph]].

\bibitem{Criado:2019tzk}
J.~C.~Criado, F.~Feruglio and S.~J.~D.~King,
JHEP \textbf{02}, 001 (2020)
[arXiv:1908.11867 [hep-ph]].

\bibitem{Wang:2019ovr}
X.~Wang and S.~Zhou,
JHEP \textbf{05}, 017 (2020)
[arXiv:1910.09473 [hep-ph]].
  
\bibitem{Zhao:2021jxg}
Y.~Zhao and H.~H.~Zhang,
JHEP \textbf{03} (2021), 002
[arXiv:2101.02266 [hep-ph]].
  
\bibitem{King:2021fhl}
S.~F.~King and Y.~L.~Zhou,
JHEP \textbf{04} (2021), 291
[arXiv:2103.02633 [hep-ph]].
  
\bibitem{Ding:2021zbg}
G.~J.~Ding, S.~F.~King and C.~Y.~Yao,
[arXiv:2103.16311 [hep-ph]].
  
\bibitem{Zhang:2021olk}
X.~Zhang and S.~Zhou,
[arXiv:2106.03433 [hep-ph]].

\bibitem{gui-jun}
Bu-Yao Qu, Xiang-Gan Liu, Ping-Tao Chen, Gui-Jun Ding
[arXiv:2106.11659 [hep-ph]].

\bibitem{Nomura:2021ewm}
T.~Nomura and H.~Okada,
[arXiv:2109.04157 [hep-ph]].



\bibitem{Novichkov:2018nkm}
P.~P.~Novichkov, J.~T.~Penedo, S.~T.~Petcov and A.~V.~Titov,
JHEP \textbf{04}, 174 (2019)
[arXiv:1812.02158 [hep-ph]].

\bibitem{Ding:2019xna}
G.~J.~Ding, S.~F.~King and X.~G.~Liu,
Phys. Rev. D \textbf{100}, no.11, 115005 (2019)
[arXiv:1903.12588 [hep-ph]].


\bibitem{Liu:2019khw}
X.~G.~Liu and G.~J.~Ding,
JHEP \textbf{08}, 134 (2019)
[arXiv:1907.01488 [hep-ph]].

\bibitem{Chen:2020udk}
P.~Chen, G.~J.~Ding, J.~N.~Lu and J.~W.~F.~Valle,
Phys. Rev. D \textbf{102}, no.9, 095014 (2020)
[arXiv:2003.02734 [hep-ph]].

\bibitem{Li:2021buv}
C.~C.~Li, X.~G.~Liu and G.~J.~Ding,
[arXiv:2108.02181 [hep-ph]].

\bibitem{Novichkov:2020eep}
P.~P.~Novichkov, J.~T.~Penedo and S.~T.~Petcov,
Nucl. Phys. B \textbf{963}, 115301 (2021)
[arXiv:2006.03058 [hep-ph]].

\bibitem{Liu:2020akv}
X.~G.~Liu, C.~Y.~Yao and G.~J.~Ding,
Phys. Rev. D \textbf{103}, no.5, 056013 (2021)
[arXiv:2006.10722 [hep-ph]].





\bibitem{Wang:2020lxk}
X.~Wang, B.~Yu and S.~Zhou,
Phys. Rev. D \textbf{103}, no.7, 076005 (2021)
[arXiv:2010.10159 [hep-ph]].

\bibitem{Yao:2020zml}
C.~Y.~Yao, X.~G.~Liu and G.~J.~Ding,
Phys. Rev. D \textbf{103}, no.9, 095013 (2021)
[arXiv:2011.03501 [hep-ph]].
  
\bibitem{Wang:2021mkw}
X.~Wang and S.~Zhou,
[arXiv:2102.04358 [hep-ph]].
  
\bibitem{Behera:2021eut}
M.~K.~Behera and R.~Mohanta,
[arXiv:2108.01059 [hep-ph]].
  


\bibitem{deMedeirosVarzielas:2019cyj}
I.~de Medeiros Varzielas, S.~F.~King and Y.~L.~Zhou,
Phys. Rev. D \textbf{101}, no.5, 055033 (2020)
[arXiv:1906.02208 [hep-ph]].

\bibitem{Kobayashi:2018bff}
T.~Kobayashi and S.~Tamba,
Phys. Rev. D \textbf{99}, no.4, 046001 (2019)
[arXiv:1811.11384 [hep-th]].

\bibitem{Kikuchi:2020nxn}
S.~Kikuchi, T.~Kobayashi, H.~Otsuka, S.~Takada and H.~Uchida,
JHEP \textbf{11}, 101 (2020)
[arXiv:2007.06188 [hep-th]].
  
\bibitem{Almumin:2021fbk}
Y.~Almumin, M.~C.~Chen, V.~Knapp-P\'erez, S.~Ramos-S\'anchez, M.~Ratz and S.~Shukla,
JHEP \textbf{05} (2021), 078
[arXiv:2102.11286 [hep-th]].

\bibitem{Ding:2021iqp}
G.~J.~Ding, F.~Feruglio and X.~G.~Liu,
SciPost Phys. \textbf{10} (2021), 133
[arXiv:2102.06716 [hep-ph]].

\bibitem{Feruglio:2021dte}
F.~Feruglio, V.~Gherardi, A.~Romanino and A.~Titov,
JHEP \textbf{05} (2021), 242
[arXiv:2101.08718 [hep-ph]].

\bibitem{Kikuchi:2021ogn}
S.~Kikuchi, T.~Kobayashi and H.~Uchida,
[arXiv:2101.00826 [hep-th]].

\bibitem{Novichkov:2021evw}
P.~P.~Novichkov, J.~T.~Penedo and S.~T.~Petcov,
JHEP \textbf{04} (2021), 206
[arXiv:2102.07488 [hep-ph]].

\bibitem{Kikuchi:2021yog}
S.~Kikuchi, T.~Kobayashi, Y.~Ogawa and H.~Uchida,
[arXiv:2112.01680 [hep-ph]].

\bibitem{Novichkov:2022wvg}
P.~P.~Novichkov, J.~T.~Penedo and S.~T.~Petcov,
[arXiv:2201.02020 [hep-ph]].


\bibitem{Altarelli:2010gt}
G.~Altarelli and F.~Feruglio,
Rev. Mod. Phys. \textbf{82}, 2701-2729 (2010)
[arXiv:1002.0211 [hep-ph]].

\bibitem{Ishimori:2010au}
H.~Ishimori, T.~Kobayashi, H.~Ohki, Y.~Shimizu, H.~Okada and M.~Tanimoto,
Prog. Theor. Phys. Suppl. \textbf{183}, 1-163 (2010)
[arXiv:1003.3552 [hep-th]].

\bibitem{Ishimori:2012zz}
H.~Ishimori, T.~Kobayashi, H.~Ohki, H.~Okada, Y.~Shimizu and M.~Tanimoto,
Lect. Notes Phys. \textbf{858}, 1-227 (2012)

\bibitem{Hernandez:2012ra}
D.~Hernandez and A.~Y.~Smirnov,
Phys. Rev. D \textbf{86}, 053014 (2012)
[arXiv:1204.0445 [hep-ph]].

\bibitem{King:2013eh}
S.~F.~King and C.~Luhn,
Rept. Prog. Phys. \textbf{76}, 056201 (2013)
[arXiv:1301.1340 [hep-ph]].

\bibitem{King:2014nza}
S.~F.~King, A.~Merle, S.~Morisi, Y.~Shimizu and M.~Tanimoto,
New J. Phys. \textbf{16}, 045018 (2014)
[arXiv:1402.4271 [hep-ph]].

\bibitem{King:2017guk}
S.~F.~King,
Prog. Part. Nucl. Phys. \textbf{94}, 217-256 (2017)
[arXiv:1701.04413 [hep-ph]].

\bibitem{Petcov:2017ggy}
S.~T.~Petcov,
Eur. Phys. J. C \textbf{78}, no.9, 709 (2018)
[arXiv:1711.10806 [hep-ph]].

\bibitem{Baur:2019iai}
A.~Baur, H.~P.~Nilles, A.~Trautner and P.~K.~S.~Vaudrevange,
Nucl. Phys. B \textbf{947}, 114737 (2019)
[arXiv:1908.00805 [hep-th]].

\bibitem{Kobayashi:2019uyt}
T.~Kobayashi, Y.~Shimizu, K.~Takagi, M.~Tanimoto, T.~H.~Tatsuishi and H.~Uchida,
Phys. Rev. D \textbf{101}, no.5, 055046 (2020)
[arXiv:1910.11553 [hep-ph]].

\bibitem{Novichkov:2019sqv}
P.~P.~Novichkov, J.~T.~Penedo, S.~T.~Petcov and A.~V.~Titov,
JHEP \textbf{07}, 165 (2019)
[arXiv:1905.11970 [hep-ph]].

\bibitem{Baur:2019kwi}
A.~Baur, H.~P.~Nilles, A.~Trautner and P.~K.~S.~Vaudrevange,
Phys. Lett. B \textbf{795}, 7-14 (2019)
[arXiv:1901.03251 [hep-th]].


\bibitem{Kobayashi:2020hoc}
T.~Kobayashi and H.~Otsuka,
Phys. Rev. D \textbf{101}, no.10, 106017 (2020)
[arXiv:2001.07972 [hep-th]].
 
 
\bibitem{Kobayashi:2020uaj}
T.~Kobayashi and H.~Otsuka,
Phys. Rev. D \textbf{102}, no.2, 026004 (2020)
[arXiv:2004.04518 [hep-th]].


\bibitem{Ishiguro:2020nuf}
K.~Ishiguro, T.~Kobayashi and H.~Otsuka,
Nucl. Phys. B \textbf{973}, 115598 (2021)
[arXiv:2010.10782 [hep-th]].


\bibitem{Tanimoto:2021ehw}
M.~Tanimoto and K.~Yamamoto,
[arXiv:2106.10919 [hep-ph]].


\bibitem{Ishiguro:2021ccl}
K.~Ishiguro, T.~Kobayashi and H.~Otsuka,
JHEP \textbf{01}, 020 (2022)
[arXiv:2107.00487 [hep-th]].



\bibitem{Chen:2019ewa}
M.~C.~Chen, S.~Ramos-S\'anchez and M.~Ratz,
Phys. Lett. B \textbf{801}, 135153 (2020)
[arXiv:1909.06910 [hep-ph]].

\bibitem{deMedeirosVarzielas:2020kji}
I.~de Medeiros Varzielas, M.~Levy and Y.~L.~Zhou,
JHEP \textbf{11}, 085 (2020)
[arXiv:2008.05329 [hep-ph]].

\bibitem{Ishiguro:2020tmo}
K.~Ishiguro, T.~Kobayashi and H.~Otsuka,
JHEP \textbf{03}, 161 (2021)
[arXiv:2011.09154 [hep-ph]].


\bibitem{Abe:2020vmv}
H.~Abe, T.~Kobayashi, S.~Uemura and J.~Yamamoto,
Phys. Rev. D \textbf{102}, no.4, 045005 (2020)
[arXiv:2003.03512 [hep-th]].

\bibitem{Novichkov:2022wvg}
P.~P.~Novichkov, J.~T.~Penedo and S.~T.~Petcov,
[arXiv:2201.02020 [hep-ph]].


\bibitem{Kikuchi:2022txy}
S.~Kikuchi, T.~Kobayashi, H.~Otsuka, M.~Tanimoto, H.~Uchida and K.~Yamamoto,
[arXiv:2201.04505 [hep-ph]].


\bibitem{Kobayashi:2021uam}
T.~Kobayashi and H.~Otsuka,
Eur. Phys. J. C \textbf{82}, no.1, 25 (2022)
[arXiv:2108.02700 [hep-ph]].



\bibitem{Kikuchi:2022bkn}
S.~Kikuchi, T.~Kobayashi, K.~Nasu, H.~Uchida and S.~Uemura,
[arXiv:2202.05425 [hep-th]].

  
\bibitem{Kajiyama:2013zla}
Y.~Kajiyama, H.~Okada and K.~Yagyu,
Nucl. Phys. B \textbf{874} (2013), 198-216
[arXiv:1303.3463 [hep-ph]].
  
\bibitem{Kajiyama:2013rla}
Y.~Kajiyama, H.~Okada and T.~Toma,
Phys. Rev. D \textbf{88} (2013) no.1, 015029
[arXiv:1303.7356 [hep-ph]].


\bibitem{Kaplunovsky:1993rd}
V.~S.~Kaplunovsky and J.~Louis,
Phys. Lett. B \textbf{306} (1993), 269-275
[arXiv:hep-th/9303040 [hep-th]].
  
\bibitem{Arkani-Hamed:2000oup}
N.~Arkani-Hamed, L.~J.~Hall, H.~Murayama, D.~Tucker-Smith and N.~Weiner,
Phys. Rev. D \textbf{64} (2001), 115011
[arXiv:hep-ph/0006312 [hep-ph]].


\bibitem{Kobayashi:2021jqu}
T.~Kobayashi, T.~Shimomura and M.~Tanimoto,
Phys. Lett. B \textbf{819}, 136452 (2021)
[arXiv:2102.10425 [hep-ph]].

  
\bibitem{Megrelidze:2016fcs}
L.~Megrelidze and Z.~Tavartkiladze,
Nucl. Phys. B \textbf{914} (2017), 553-576
[arXiv:1609.07344 [hep-ph]].
  
\bibitem{Aghanim:2018eyx} 
  N.~Aghanim {\it et al.} [Planck Collaboration],
  arXiv:1807.06209 [astro-ph.CO].

  \bibitem{PDG} M. Tanabashi et al. (Particle Data Group), Phys. Rev. D {\bf 98}, 030001 (2018). 

\bibitem{KamLAND-Zen:2016pfg} 
  A.~Gando {\it et al.} [KamLAND-Zen Collaboration],
  Phys.\ Rev.\ Lett.\  {\bf 117}, no. 8, 082503 (2016)
  Addendum: [Phys.\ Rev.\ Lett.\  {\bf 117}, no. 10, 109903 (2016)]
  [arXiv:1605.02889 [hep-ex]].

\bibitem{Esteban:2018azc} 
  I.~Esteban, M.~C.~Gonzalez-Garcia, A.~Hernandez-Cabezudo, M.~Maltoni and T.~Schwetz,
  JHEP {\bf 1901}, 106 (2019)
  [arXiv:1811.05487 [hep-ph]].
  
\bibitem{Hambye:2009pw}
T.~Hambye, F.~S.~Ling, L.~Lopez Honorez and J.~Rocher,
JHEP \textbf{07} (2009), 090
[erratum: JHEP \textbf{05} (2010), 066]
[arXiv:0903.4010 [hep-ph]].

\bibitem{Kanemura:2010sh} 
  S.~Kanemura, S.~Matsumoto, T.~Nabeshima and N.~Okada,
  Phys.\ Rev.\ D {\bf 82}, 055026 (2010)
  [arXiv:1005.5651 [hep-ph]].
  \end{thebibliography}
\end{document}